\documentclass[aps,prd,nofootinbib,twocolumn,superscriptaddress,floatfix,notitlepage,10pt]{revtex4-1}
\usepackage[export]{adjustbox} 
\usepackage[utf8]{inputenc}
\usepackage[dvipsnames]{xcolor}
\usepackage{graphicx}
\usepackage{amsmath,amsfonts,amssymb}
\usepackage[breaklinks,colorlinks,urlcolor=Plum,citecolor=MidnightBlue,linkcolor=WildStrawberry]{hyperref}
\newcommand{\MPl}{M_{\textrm{\tiny{Pl}}}}
\usepackage{bm}
\usepackage{acronym}
\usepackage{xspace}
\usepackage{multirow}

\usepackage{tabularray}
\usepackage{tabularx}
\usepackage{orcidlink}
\usepackage{ragged2e}
\usepackage{setspace}
\usepackage{comment}

\usepackage{listings}
\usepackage[ruled,vlined]{algorithm2e}

\SetCommentSty{mycommfont}
\definecolor{forestgreen}{rgb}{0.13,0.35,0.13}

\newacro{S/N}{signal-to-noise ratio}

\newacro{PN}{post-Newtonian}

\newacro{O3}{the third observing run}

\newacro{O2}{the second observing run}

\newacro{CW}{\emph{Continuous Wave}}

\newcommand{\bea}{\begin{equation}\begin{aligned}} 
\newcommand{\eea}{\end{aligned}\end{equation}}
\newcommand{\be}{\begin{equation}}
\newcommand{\ee}{\end{equation}}

\newcommand{\msun}{M_{\odot}}

\newcommand{\td}{{\rm d}}
\newcommand{\pd}{\partial}

\newcommand{\eps}{\epsilon}

\definecolor{rossocorsa}{rgb}{0.83, 0.0, 0.0}

\begin{document}

\title{\texorpdfstring{Beware of the running $n_s$ when producing heavy primordial black holes}{Beware of the running ns when producing heavy primordial black holes}}

\author{Sasha Allegrini\orcidlink{0009-0004-2664-7440}}
\thanks{ \href{mailto:xxx}{sasha.allegrini@kbfi.ee}}
\affiliation{Keemilise ja Bioloogilise F\"u\"usika Instituut, R\"avala pst. 10, 10143 Tallinn, Estonia}
\affiliation{Tallinn University of Technology, Akadeemia tee 23, Tallinn, 12618, Estonia}

\author{Antonio J. Iovino\orcidlink{0000-0002-8531-5962}}
\thanks{ \href{mailto:a.iovino@nyu.edu}{a.iovino@nyu.edu}}
\affiliation{Center for Astrophysics and Space Science (CASS), New York University Abu Dhabi, PO Box 129188, Abu Dhabi, UAE}
\author{Hardi~Veerm\"ae\orcidlink{0000-0003-1845-1355}}
\email{hardi.veermae@cern.ch}
\affiliation{Keemilise ja Bioloogilise F\"u\"usika Instituut, R\"avala pst. 10, 10143 Tallinn, Estonia}

\date{\today}
\begin{abstract}
We examine single-field inflationary models for the formation of primordial black holes (PBHs). By analyzing the latest observations from the Atacama Cosmology Telescope (ACT)~\cite{ACT:2025fju, ACT:2025tim}, we demonstrate that the observed preference for positive running ($\alpha_s$) of the scalar spectral index $n_s$ imposes significant restrictions on the parameter space for a large class of ultra-slow-roll (USR) models. This tension becomes progressively pronounced for more massive PBHs, posing substantial challenges for USR models to yield a detectable PBH abundance, especially in the mass range probed by ongoing and future gravitational-wave experiments such as LIGO-Virgo-KAGRA and the Einstein Telescope. However, this discrepancy is minimal for asteroid-mass PBHs, which are still capable of feasibly constituting the entirety of dark matter. To numerically probe the six-dimensional parameter space of the models, we adapted a Markov Chain Monte Carlo approach to efficiently scan over the viable configurations. Our results further indicate that, in non-minimally coupled polynomial inflation, a viable cosmic microwave background (CMB) spectrum is best obtained at an inflection point for which second-order slow-roll approximation is necessary for precise CMB predictions.
\end{abstract}

\maketitle

%-------------------------------------------------------------------------------
\section{Introduction}
\label{sec:Introduction}
\begin{figure}[!t]
    \centering
    \includegraphics[width=0.495\textwidth]{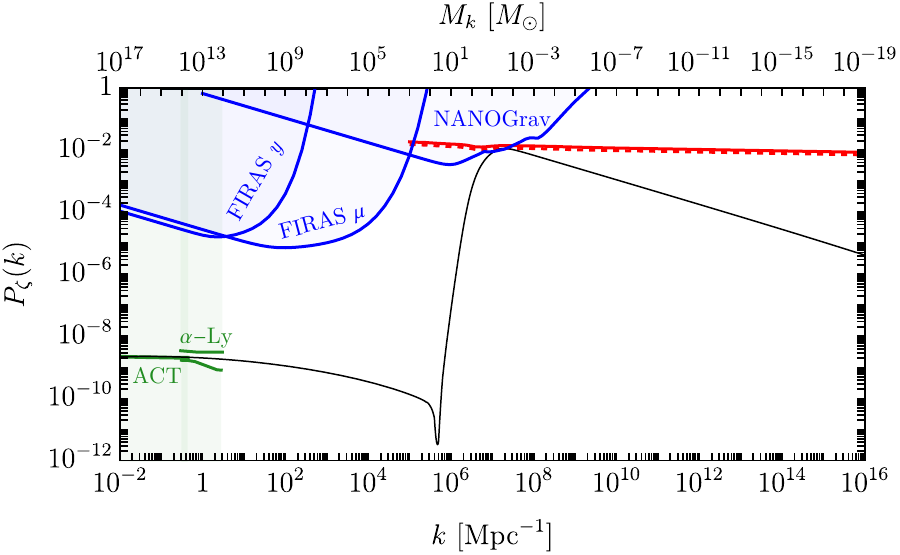}
    \caption{ \justifying \it 
    Existing constraints (colored lines) on the primordial curvature power spectrum as a function of the scale $k$, assuming negligible primordial non-Gaussianities. While the blue constraints on the amplitude refer to the amplitude evaluated at the position of the main peak $k_{\rm pk}$, the green constraints apply to the entire spectrum. The black line corresponds to a benchmark power spectrum. The red lines show the amplitude corresponding to $f_{\rm PBH}=1$ for the TS (solid) and PT (Dashed) formalism (see Sec.\,\ref{subsec:ABU} for more details).  
    }
\label{fig:PS}
\end{figure}

%-------------------------------------------------------------------------------
Precision measurements of the cosmic microwave background (CMB) provide strong evidence for inflation as the predominant framework to explain the observed flatness and homogeneity of the universe~\cite{Starobinsky:1980te, Guth:1980zm, Linde:1981mu, Albrecht:1982wi}. The inflationary epoch could further account for dark matter (DM) in the form of primordial black holes (PBHs), provided that sufficiently large curvature perturbations are generated during this period~\cite{Zeldovich:1967lct, Hawking:1974rv, Chapline:1975ojl, Carr:1975qj}. However, since the amplitude of perturbations at the CMB scales $P_\zeta(k_{\rm CMB}) \approx \mathcal{O}(10^{-9})$ is significantly smaller than that required for PBH production $P_\zeta(k_{\rm PBH}) \approx \mathcal{O}(10^{-2})$, the curvature perturbations must be significantly enhanced at smaller scales as inflation progresses. An example of such a spectrum is shown in Fig.~\ref{fig:PS}. In this paper, we will focus on one of the simplest inflationary scenarios in which such an enhancement can be realized: single-field models with a brief phase of ultra slow roll (USR)~\cite{Ivanov:1994pa, Leach:2000ea, Bugaev:2008gw, Alabidi:2009bk, Drees:2011hb, Drees:2011yz, Alabidi:2012ex, Kannike:2017bxn, Garcia-Bellido:2017mdw, Ballesteros:2017fsr, Di:2017ndc, Germani:2017bcs, Cicoli:2018asa, Ozsoy:2018flq, Bhaumik:2019tvl, Ballesteros:2020qam, Karam:2022nym, Franciolini:2022pav, Balaji:2022rsy, Frolovsky:2023hqd, Allegrini:2024ooy, Briaud:2025hra}. 

We aim to study USR models and their compatibility with current observational constraints on CMB observables, such as the tensor-to-scalar ratio $r$, the scalar spectral index $n_s$, and its running $\alpha_s$. At scales relevant to the CMB, the primordial curvature power spectrum is nearly scale invariant and can be captured by an expansion around some pivot scale $k_{\star}$, 
\be\label{eq:ParametricPS}
    P_{\zeta}(k) 
    = A_s\left(\frac{k\,}{k_{\star}}\right)^{{n_s - 1 + \frac{\alpha_s}{2}\ln\frac{k\,}{k_\star} + \dots}}\,,
\ee
where $A_s$ is the amplitude of the power spectrum at $k_{\star}$, $n_s$ is the scalar spectral index and $\alpha_s \equiv \td n_s/\td\log k$ characterizes the running of the spectral index. Combining the data of ACT~\cite{ACT:2025fju, ACT:2025tim} with the year 1 data from DESI~\cite{DESI:2024uvr, DESI:2024mwx}, gives a spectral index 
\be
    n_s = 0.9743 \pm 0.0034 \quad \mbox{(ACT+Planck)}
\ee
at $k_{\rm CMB} \equiv 0.05 \,{\rm Mpc}^{-1}$ with the uncertainty reported at the $1\sigma$ level. This represents a notable shift towards a scale-invariant spectrum compared to the earlier measurement $n_s = 0.9641 \pm 0.0044$ by Planck~\cite{Planck:2018vyg}. 

Although no significant evidence has been found for the running of the spectral index, the bounds on it have been modified by the ACT observations~\cite{ACT:2025tim} 
\be
    \alpha_s = 0.0062 \pm 0.0052 \quad \mbox{(ACT+Planck)}
\ee
showing a preference for positive running in contrast to the combined fits to Planck and the Lyman-$\alpha$ forest that indicated a preference for a negative $\alpha_s = -0.010 \pm 0.004$~\cite{Palanque-Delabrouille:2019iyz}.
Similarly, the existing upper bound on the tensor-to-scalar ratio is slightly more stringent, i.e., from $r \lesssim 0.08$ to $r \lesssim 0.05$ at the $2\sigma$ level.

The consistency of USR models with cosmological observations has been studied extensively. Typically, inflation can be divided into a period responsible for the CMB spectrum and a subsequent period of constant-$\nu$ inflation, consisting of a USR phase and a subsequent dual constant-roll (CR) phase, during which the spectral peak is generated~\cite{Kannike:2017bxn, Karam:2022nym, Briaud:2025hra}. To produce sufficiently heavy PBHs that would not have evaporated by today, the initial slow-roll (SR) period should not last longer than $37$ $e$-folds -- thus, roughly speaking, this implies that one needs to consider potentials that generate the spectral characteristics observed in CMB in less than $37$ $e$-folds instead of the usual $50-60$ $e$-folds. The duration of the second phase can be chosen so that a total of $50-60$ $e$-folds of inflation is obtained. Typically, the field excursion in the second phase is small, as the inflaton spends the second phase crossing a flat feature or a local maximum. 

The reduced duration of the initial SR phase, combined with the presence of a spectral peak high enough to trigger PBH production, can pose a notable challenge for model building. In fact, both requirements tend to move the spectral features away from scale invariance, which can cause conflicts between PBH scenarios and CMB observations, especially when PBHs are heavy~\cite{Kannike:2017bxn, Ballesteros:2020qam}. Recent ACT measurements have reinforced this issue~\cite{Frolovsky:2025iao,Merchand:2025bzt}. In particular, the higher values of $n_s$ implied by ACT exclude some of the non-polynomial USR models commonly studied, such as fibre~\cite{Cicoli:2018asa} and Higgs inflation~\cite{Garcia-Bellido:2017mdw,Germani:2017bcs} featuring an inflection point, are now disfavored by the new data. 

The goal of this work is to quantify how well single-field inflation can simultaneously generate PBHs and account for CMB observations. To this end, we present a simple algorithm for constructing viable models for PBHs by introducing features by hand. Additionally, we will consider models in the class of non-minimally coupled polynomial inflation and perform a scan over the parameter space to identify PBH models consistent with CMB measurements.
As we demonstrate in this work, USR models tend to predict negative running of $n_s$, making them increasingly constrained by recent ACT data, which favour positive running. These updated bounds play a crucial role in excluding large regions of parameter space for such models, especially for heavy, i.e. solarish, PBHs.

Recently, the SPT collaboration~\cite{SPT-3G:2025bzu} has found a measurement of $n_s$ in perfect agreement with the previous PLANCK data set, i.e., $n_s=0.9684 \pm 0.0030$. Furthermore, it has been pointed out that, due to the tension between the DESI BAO and CMB data, caution must be taken when interpreting the measurements of inflationary observables such as $n_s$~\cite{Ferreira:2025lrd}. %[This discrepancy may be attributable an underestimate of the optical depth~\cite{Giare:2023ejv, Allali:2025yvp}.] 
As the resolution of this tension lies beyond the scope of this paper, we will consider both the results of Planck~\cite{Planck:2018vyg} and ACT+Planck~\cite{ACT:2025tim} when constructing viable models. This approach enables us to draw fairly robust conclusions that are independent of the precise dataset we are taking into account.

The paper is structured as follows. In Sec.~\ref{sec:Basic}, we introduce some basic concepts underlying the formation of PBHs from single-field inflation. We then describe models capable of generating a USR phase---starting with a toy model, followed by an analysis of two polynomial inflationary potentials---in Sec.~\ref{sec:modelbulding}. In Sec.~\ref{sec:fit}, we address the central question of this work: how the recent ACT dataset constrains the parameter space for PBH production in single-field models. Finally, we conclude in Sec.~\ref{sec:conclusions}.

\section{PBHs from single-field inflation} 
\label{sec:Basic}
%-------------------------------------------------------------------------------

The most general action that describes a non-minimally coupled scalar takes the form
\be
S\!=\! \int\td^4 x \sqrt{\!-\!g} \left[\!-\frac{\MPl^2}{2} \Omega(\phi) R \!+\! \frac{K(\phi)}{2} (\partial \phi)^2 \!-\! V(\phi) \right],
    \label{eq:action}
\ee
where $\Omega(\phi)$, $K(\phi)$ and $V(\phi)$ denote the non-minimal coupling, the coefficient of the non-canonical kinetic term, and the Jordan frame scalar potential, respectively.  The theory admits an Einstein frame formulation after the Weyl transformation $g_{\mu\nu} \to \bar g_{\mu\nu} = g_{\mu\nu}/\Omega(\phi)$, so that
\be
    S
    \!=\! \int\td^4 x \sqrt{\!-\!\bar{g}} \left[\!-\frac{\MPl^2}{2} \bar{R} \!+\! \frac{\bar{K}(\phi)}{2} (\partial \phi)^2 \!-\! \bar{V}(\phi) \right],
\ee
where the barred quantities, corresponding to the Einstein frame, read
\be\label{eq:barV,barK}
    \bar{V}(\phi) = \frac{V}{\Omega^2}\,, \qquad
    \bar{K} = \frac{K}{\Omega} + \frac{3 \MPl^2}{2} \left(\frac{\pd_{\phi}\Omega}{\Omega} \right)^2\,.
\ee
The non-canonical kinetic term can always be absorbed into the potential with the field redefinition $\phi \to \bar \phi$ defined by $\td \bar \phi = \sqrt{\bar K(\phi)} \td \phi$, so any single field model can be fully characterized by the Einstein frame potential $\bar{V}(\bar \phi) \equiv \bar{V}(\phi(\bar \phi))$. However, since the mapping $\phi \to \bar \phi$ cannot be found analytically in most cases, we will mainly work with the Jordan frame field $\phi$ in numerical estimates.

The evolution of the homogeneous inflaton in a Friedmann-Lema\^itre-Robertson-Walker (FLRW) background is dictated by the Klein-Gordon equation
$\ddot{\bar \phi} + 3H\dot {\bar \phi} + \pd_{\bar \phi}\bar V = 0$ and the Friedmann equation $3 \MPl^2H^2 = \rho_\phi$, where $\rho_\phi = \dot {\bar \phi}^2/2+\bar V$ and $H$ is the Hubble parameter. Using the number of $e$-folds $N$ as the new time variable, these equations can be combined into a system of two first order equations, which, expressed in terms of the Jordan frame field, take the form
\bea\label{eq:eom_phi,y}
    \pd_N y &= -\left(3 - \frac{y^2}{2}\right) \left( y + \frac{\MPl}{\sqrt{\bar K}(\phi)} \pd_\phi \ln \bar V(\phi) \right) \,,
    \\
    \pd_N \phi &= \MPl y / \sqrt{\bar{K}(\phi)}\,.
\eea
Note that $y = \pd_N \bar \phi/\MPl$ is simply the derivative of the Einstein frame field. In addition, the overall scale of the potential has dropped out from the equations of motion and appears only in the Hubble parameter, $H^2 = \bar V(\phi)/(\MPl^2(3-y^2/2))$.

\subsection{Primordial Power Spectrum} \label{sec:IIA}

The leading order evolution of curvature perturbations $\zeta$ is governed by the Mukhanov–Sasaki equation
\be \label{eq:MukSas}
    \frac{\pd^2 u_k}{\pd \tau^2} + \left(k^2 - \frac{1}{z}\frac{\pd^2 z}{\pd \tau^2}\right) u_k = 0\,,
\ee
where  $u_k = z \zeta_k$ is the Mukhanov–Sasaki variable, $z \equiv a \dot {\bar \phi}/H = a y \MPl$, $a \equiv e^N$ is the scale factor and derivatives are taken in conformal time $\tau$. The initial conditions for the mode functions are set assuming that all modes follow the Bunch–Davies vacuum deep inside the horizon~\cite{Bunch:1978yq}, that is, when $N \ll N_k$, where $N_k$ is the time of horizon crossing defined by $k = a(N_k) H(N_k)$. The dimensionless scalar curvature power spectrum is then given by
\bea
\label{eq:PS}
    P_{\zeta}(k) 
    = \frac{k^3}{2\pi^2} \left| \frac{u_k}{z} \right|^2_{N \gg N_k}\,.
\eea
Curvature perturbations $\zeta_k=u_k/z$ generally freeze after horizon exit ($N\gg N_k$), since $u_k\propto k$ when $k^2\ll|\partial_\tau^2 z/z|\sim\mathcal O((aH)^2)$. However, PBH-producing models often exhibit superhorizon enhancement during the USR phase~\cite{Leach:2001zf}, so one must follow the superhorizon evolution of $\zeta_k$ for $N>N_k$ at least until the end of the USR stage. However, although modes that exit after the USR phase are not subject to superhorizon enhancement, the transition from the SR phase to the USR phase can strongly affect the subhorizon evolution and leave an imprint on the spectral peak in the form of spectral modulation~\cite{Briaud:2025hra}. The Bunch-Davies vacuum can therefore be imposed only before the transition into USR.

It is convenient to introduce the Hubble flow parameters defined recursively as $\eps_{i+1} = \td \ln \eps_i/\td N$, with $\eps_0 = 1/H$, so that $\eps_1 = y^2/2$, $\eps_2 = 2 \pd_N y/y$. The time-dependence in the Mukhanov–Sasaki equation can then be captured via the parameter $\nu$, defined via
\be
    \frac{1}{z}\frac{\pd^2 z}{\pd \tau^2} \equiv (aH)^2 \left(\nu^2 - \frac{1}{4} \right)\,,
\ee
which, expressed via the Hubble-flow parameters, takes the form
\be\label{eq:nu}
    \nu^2 
    = \frac{9}{4} - \eps_1 + \frac{3}{2}\eps_2 - \frac{1}{2}\eps_1\eps_2 + \frac{1}{4}\eps_2^2 + \frac{1}{2}\eps_2\eps_3\,.
\ee
Typical single-field models for PBHs featuring shallow local minima induce a phase of constant-$\nu$ inflation during which a sizable peak in the power spectrum is created~\cite{Karam:2022nym, Briaud:2025hra}. This phase consists of a USR phase\footnote{As in Ref.~\cite{Ballesteros:2020qam}, we define the USR phase by the condition $\eps_2 < -3$ instead of the narrower definition $\eps_2 = - 6$.} with $\eps_{2,\rm USR} \lesssim -6$ in which the inflaton decelerates and is followed by a CR (or SR) phase with $\eps_{2,\rm CR} \gtrsim 0$. Since $\nu = const.$ as the inflation passes through the USR to CR phases, the Mukhanov–Sasaki equation is not affected by the crossing from USR to CR and therefore the deceleration of the inflaton during USR and its subsequent satisfy the Wands dual relation $\eps_{2,\rm USR} + \eps_{2,\rm CR}  = -6$~\cite{Wands:1998yp}. This effectively yields a broken power law shape for the peak in the power spectrum, as long as the transition to USR is sufficiently smooth~\cite{Cole:2022xqc, Karam:2022nym, Franciolini:2022pav, Briaud:2025hra}.

The primordial tensor power spectrum is
\begin{align}
    P_T(k) = 
    \frac{4k^3}{\pi^2}\left|
    \frac{v_k}{a\MPl}
    \right|^2_{N \gg N_k}\,,
\end{align}
where the tensor modes $v_k/a$ evolve according to
\be\label{eq:eom_tensor}
    \frac{\pd^2 v_k}{\pd \tau^2} + \left(k^2 - \frac{1}{a}\frac{\pd^2 a}{\pd \tau^2}\right) v_k = 0\,,
\ee
with initial conditions set by the Bunch-Davies vacuum. In terms of the Hubble-flow parameters, the time dependence of the frequency is
\be
    \frac{1}{a}\frac{\pd^2 a}{\pd \tau^2} = (aH)^2(1-\eps_1)\,.
\ee
Since $\eps_1 \equiv - \pd_N \ln H$ depends only on the first derivative of $H$, it is less sensitive to sharp changes in the background evolution. In particular, unlike $\nu$, it does not contain peaked features at the CMB scales. We checked that the SR estimate of $r$, discussed in detail in appendix~\ref{app:SRandCR}, is in excellent agreement with the numerical solution of~\eqref{eq:eom_tensor} for all models considered.

\subsection{PBH abundance}
\label{subsec:ABU}
Large curvature fluctuations can collapse to PBHs after horizon reentry with their mass following the critical scaling law~\cite{Choptuik:1992jv, Niemeyer:1997mt, Niemeyer:1999ak}
\be
    M_{\rm PBH}(\mathcal{C}) 
    = \mathcal{K} M_k (\mathcal{C}-\mathcal{C}_{\rm th})^{\gamma}\,,
\ee
where $\mathcal{C}$ is the compaction and a black hole is formed only it exceeds the threshold $\mathcal{C}_{\rm th}$ that depends on the shape of the power spectrum~\cite{Musco:2020jjb}, $\gamma = 0.38$ is a critical exponent, $\mathcal{K} = 4.4$ is a coefficient, and
\be\label{eq:HorizonMasskH}
    M_k(t) 
    \simeq 14 \msun \times \left[\frac{k}{10^{6}{\rm Mpc}^{-1}} \right]^{-2}
\ee
is the horizon mass at the time of the re-entry of the mode $k$. We also include the effect of the QCD phase transition on these parameters~\cite{Musco:2023dak}.
At the time of re-entry of $k$, the fraction of radiation $\beta_k(M_{\rm PBH}) \td  \ln M_{\rm PBH}$ collapsing into PBHs of mass $M_{\rm PBH}$ is
\be\label{eq:betak}
    \beta_k(M_{\rm PBH})
    = \int_{\mathcal{C}_{\rm th}} \! \td\mathcal{C} \, P_k(\mathcal{C}) \frac{M_{\rm PBH}}{M_k}  \delta\left[ \ln\frac{M_{\rm PBH}}{M_{\rm PBH}(\mathcal{C})} \right]\!,
\ee
where $P_k(\mathcal{C})$ denotes the probability distribution of the compaction. The resulting present-day PBH mass function reads~\footnote{It is normalized so that the total abundance of dark matter in the form of PBHs is
\be
   f_{\rm PBH} 
   = \int \td \ln M_{\rm PBH} \, f_{\rm PBH}(M_{\rm PBH})\,.
\ee}
\be\label{eq:df_PBH}
    f_{\rm PBH}(M_{\rm PBH})
= \frac{1}{\Omega_{\rm DM}}\int \frac{\td M_k}{M_k} \, \beta_k(M_{\rm PBH} ) \left(\frac{M_{\rm eq}}{M_k}\right)^{1/2} \!\!,
\ee
where $M_{\rm eq} \approx 2.8\times 10^{17}\,\,M_{\odot}$ is the horizon mass at the time of matter-radiation equality and $\Omega_{\rm  DM} = 0.12h^{-2}$ is the cold dark matter density~\cite{Planck:2018jri}.

Estimation of the abundance and mass distribution of PBHs remains an open problem, as current methods rely on approximations. The two most common approaches in the literature rely on threshold statistics of the compaction function (TS) and peaks theory (PT), both of which can be recast in the form~\eqref{eq:betak} but with different choices of $P_k(\mathcal{C})$~\cite {Iovino:2024tyg}. We also remark that, since primordial non-Gaussianities tend to be relatively insignificant in common USR models~\cite{Atal:2018neu, Firouzjahi:2023xke, Frosina:2023nxu, Ballesteros:2024pbe}, we will neglect their impact in the computation of the abundance~\cite{Young:2022phe, Ferrante:2022mui, Gow:2022jfb} and on the COBE Far Infrared Absolute Spectrophotometer (FIRAS) constraints~\cite{Sharma:2024img, Byrnes:2024vjt, Pritchard:2025yda} shown, e.g., in Figs.~\ref{fig:PS} and~\ref{fig:Toy}. 

\subsection{Scalar-induced gravitational waves}

At smaller scales compared to the CMB, the amplitude of the curvature power spectrum is constrained by gravitational waves experiments, i.e. NANOGrav in Fig.~\ref{fig:PS}. Indeed, the enhanced scalar perturbations that can give rise to a PBH population will also emit tensor modes because of second-order effects around the epoch of horizon crossing\,\cite{Matarrese:1997ay,  Matarrese:1992rp,Carbone:2004iv, Domenech:2020kqm, Bruni:1996im,Iovino:2025xkq}. The resulting gravitational wave background, which is referred to as Scalar-Induced Gravitational Waves (SIGW), possesses a spectrum 
\be\label{eq:GW}
h^2 \Omega_{\rm GW} (k) =
    \frac{h^2 \Omega_r }{24}
    \frac{g_*}{g_*^0}
    \left(\frac{g_{*s}}{g_{*s}^0}\right)^{-\frac{4}{3}}
    {\cal P}_h (k),
\ee
where $g_{*s} \equiv g_{*s} \left( T_k\right)$ and $g_* \equiv g_* \left( T_k\right)$ are the effective entropy and energy degrees of freedom (evaluated at the time of horizon crossing of mode $k$ and at present with the superscript $0$), 
while $h^2\Omega_r = 4.2\times  10^{-5}$ is the current radiation abundance. This expression assumes that the SIGW was generated during radiation domination.
The tensor mode power spectrum is~\cite{Kohri:2018awv,Espinosa:2018eve}
\bea
\label{eq:P_h_ts}
     \mathcal{P}_h (k)
     = &
    4 \int_1^\infty \td t \int_{0}^{1}\td s 
    \left [ \frac{(t^2-1)(1-s^2)}{t^2-s^2} \right ]^2 
    \\ 
    & \qquad \quad \times 
    {\cal I}_{t,s}^2 \,
     \mathcal{P}_\zeta\left(k\frac{t-s}{2}\right) 
     \mathcal{P}_\zeta\left(k\frac{t+s}{2}\right),  
\eea
where 
\bea
\label{I_RD_osc_ave_ts}
    {\cal I}_{t,s}^2 
    & =
    \frac{288(s^2\!+\!t^2\!-\!6)^2}{(t^2-s^2)^6}
    %\nonumber\\ \times & 
    \Bigg [ \frac{\pi^2}{4} (s^2\!+\!t^2\!-\!6)^2 \,\Theta (t\!-\!\sqrt{3})  
    \\
    & +
    \left(t^2 - s^2 - \frac{1}{2} (s^2+t^2-6) \log \left| \frac{t^2-3}{3-s^2} \right| 
    \right) ^2 
    \Bigg ].
\eea
is a transfer function.
Here, we neglect corrections to the SIGW background arising from primordial non-Gaussianities~\cite{Cai:2018dig,Unal:2018yaa,Yuan:2020iwf,Atal:2021jyo,Adshead:2021hnm,Abe:2022xur,Chang:2022nzu,Garcia-Saenz:2022tzu,Li:2023qua,Perna:2024ehx,Iovino:2024sgs,Zeng:2025cer,Li:2025met} and from alternative cosmic histories~\cite{Dalianis:2019asr,Bhattacharya:2019bvk,Bhattacharya:2020lhc,Papanikolaou:2022chm,Papanikolaou:2022cvo,Ireland:2023avg,Ghoshal:2023sfa,Papanikolaou:2024kjb,Domenech:2024rks,Yogesh:2025hll}, assuming instead a perfectly radiation-dominated universe. We also ignore small modifications due to variations in the sound speed during the QCD era (see, e.g.,~\cite{Hajkarim:2019nbx,Abe:2020sqb,Franciolini:2023wjm}).
Requiring that the emitted signal does not overproduce the signal registered by the NANOGrav collaboration~\cite{NANOGrav:2023gor,NANOGrav:2023hvm}, allow us to put constraints on the amplitude of the power spectrum~\cite{Iovino:2024tyg}.

The constraints on the curvature power spectrum over the entire range of scales covered by the inflationary dynamics are depicted in Fig.~\ref{fig:PS}, together with a benchmark power spectrum associated with sub-solar mass PBHs.  
We plot the region excluded by ACT measurements, Ref.~\cite{ACT:2025tim}, the FIRAS bound on CMB spectral distortions, Ref.~\cite{Chluba:2012we} (see also Ref.~\cite{Jeong:2014gna, Sharma:2024img, Byrnes:2024vjt, Iovino:2024tyg}), the bound obtained from Lyman-$\alpha$ forest data~\cite{Bird:2010mp} and the bound from NANOGRAV~\cite{NANOGrav:2023gor, NANOGrav:2023hde}. For the NANOGRAV and FIRAS experiments, we account for the shape dependence of the power spectrum in the computation of the constraints, following closely Ref.~\cite{Iovino:2024tyg}\footnote{Including an astrophysical foreground, the NANOGrav constraints become weaker~\cite{Cecchini:2025oks}.}.
In the same figure, we show the required amplitude of the power spectrum to get $f_{\rm PBH}=1$ for the TS approach (Red-Solid line) and the PT approach (Red-Dashed line). We see that the TS approach tends to result in a smaller abundance than PT in agreement with earlier studies~\cite{Green:2004wb, Young:2014ana, DeLuca:2019qsy, Iovino:2024tyg, Pi:2024ert})\footnote{Recently in Ref.~\cite{Raatikainen:2025gpd} the authors have shown that stochastic kicks can increase the PBH abundance in USR models. As we are interested only in the order-of-magnitude of the power spectrum amplitude more than a precise computation of the abundance, such effects are beyond the scope of this work. Similar recent studies demonstrate the fast evolution in PBH abundance computation (see also~\cite{Ianniccari:2024bkh,Pi:2024ert, Fumagalli:2024kxe}).}.

%-------------------------------------------------------------------------------
\section{Quick guide for model building} 
\label{sec:modelbulding}
%-------------------------------------------------------------------------------

In this section, we introduce the tools for model building that lead to a long enough inflation, with inflationary parameters compatible with observational constraints, and at the same time, a power spectrum with the amplitude enhanced in some range of scales, capable of generating a sizable population of PBHs. The general outline of the approach goes as follows:
\begin{itemize}
    \item Take a potential that reproduces the CMB observations with $N_{\rm SR}$ $e$-folds of SR.
    
    \item Introduce a localized feature that induces a USR+CR and slows inflation by an additional $N_2$ $e$-folds. The spectral peak can be generated during that phase.
    
    \item Inflation can return to a final SR stage of $N_3$ $e$-folds. We indicate the total number of $e$-folds during the SR regime as 
    \be
        N_{\rm SR} \equiv N_1+N_3
    \ee
    and the total duration of inflation as 
    \be
        N_{\rm tot}  \equiv N_{\rm SR}+N_2
    \ee
\end{itemize}
The inflationary observables are then controlled by $N_{\rm SR}$. 
The PBH mass is controlled by $N_1$, as it determines the ratio of the horizon masses corresponding to CMB scales and the scales that exit during the USR phase. The peak of the power spectrum, $k_{\rm pk}$, the horizon mass, and the duration of the first SR phase, $N_1$, are approximately related as
\bea\label{eq:kN1}
    k_{\rm pk} &\simeq 0.05\,\textrm{Mpc}^{-1} \times e^{N_1} \,,
    \\
    M_{k} &\simeq 1\,\msun \times e^{-2(N_1-18)} \,.
\eea
Thus, one should select the range of 
\be
     N_1 \in (17,37)
\ee
$e$-folds to produce non-evaporating PBHs in the mass range $M_{\rm PBH} \simeq \left[10^{-17},10\right]\,\msun$, where $f_{\rm PBH} \gtrsim 10^{-5}$ is permitted~\cite{Carr:2026hot}. $N_1$ must be larger in models for evaporating PBHs, and smaller in models for PBH seeds of supermassive black holes~\cite{Carr:2018rid, Vaskonen:2020lbd, Serpico:2020ehh}.

Setting a sufficiently large enhancement of the power spectrum at the relevant scales, one can derive a lower bound on the duration of the second phase, $N_2 \approx N_{\rm USR} + N_{\rm CR}$. To estimate this range, we make use of Wands duality~\cite{Wands:1998yp, Karam:2022nym}, according to which the parameter $\nu$ remains constant during the transition from the USR to the CR phase~\cite{Karam:2022nym}. Assuming that the height of the spectrum before the peak is roughly at the same level as it was during CMB $P_{\zeta,\rm PBH} \approx 10^{-9}$, and at the peak we have $P_{\zeta,\rm PBH} \approx 10^{-2}$, the enhancement required during the USR phase is roughly~\cite{Karam:2022nym}
\be
    e^{(3+2\nu)N_{\rm USR}} = \frac{P_{\zeta,\rm PBH}}{P_{\zeta,\rm CMB}} \simeq 10^{7}.
\ee
The CR phase must then last sufficiently long for the power spectrum to relax back to its CMB value. Assuming a power-law evolution ($\nu \approx const.$), this implies~\cite{Karam:2022nym}
\be
     e^{(3-2\nu)N_{\rm CR}} = \frac{P_{\zeta,\rm CMB}}{P_{\zeta,\rm PBH}} \simeq 10^{-7}.
\ee
However, since $\nu$ can increase toward the end of the CR phase (see Fig.~\ref{fig:evo}), the last expression provides an \emph{upper bound} on $N_2$ instead, while the estimate on the duration of the USR phase alone gives a lower bound. Altogether we can estimate that
\be
     \frac{16.1}{3+2\nu} \lesssim N_2 \lesssim \frac{16.1 \times 4\nu}{4\nu^2 - 9}
\ee
When the last phase is in SR, that is $\nu \approx 3/2$, then the upper bound diverges and we only have $ N_2 \gtrsim N_{\rm USR} \approx 2.7$, which agrees with the expectation of $\mathcal{O}(3)$ $e$-folds of USR in typical inflationary models for PBHs. In the $\nu \approx 3/2$ case, however, $N_2$ is expected to be determined mostly by the duration of the subsequent SR phase, which is typically much longer than the USR phase proceeding it. Finally, in the limit $\nu \to \infty$, one finds that $N_2 \to 0$, implying that the bounds on $N_2$ depends on the largest physically viable value of $\nu$ in the specific model.

\subsection{A bumpy toy model}
\label{sec:toy_model}

We begin by considering a logarithmic inflationary potential with an embedded localized feature responsible for enhancing the scalar power spectrum, 
\be
\label{eq:V_log+bump}
    V(\phi) 
    = V_0 \left(1 + \beta \ln\left(\frac{\phi}{\MPl}\right)\right) + V_{\rm bump}(\phi)\,.
\ee
Logarithmic inflationary potentials have been previously studied in the literature, particularly in the context of radiative corrections to otherwise flat tree-level potentials, especially in the context of Loop inflation~\cite{Dvali:1994ms,Binetruy:1996xj,Halyo:1996pp} and supersymmetric models~\cite{Espinosa:1998ks,Jeannerot:2005mc,Rocher:2004et} (for other similar models see the review~\cite{Martin:2013tda}). These models typically exhibit a slow logarithmic roll that can sustain inflation while introducing a mild scale dependence in the scalar power spectrum. The simplicity of the logarithmic form makes it a useful starting point for constructing analytically tractable models with phenomenological flexibility. To obtain a USR phase, we add a bump by hand.
Although ad hoc in nature, the addition of such features provides a simple and effective method for building toy models for producing PBHs~\cite{Mishra:2019pzq,Bhatt:2022mmn}. As mentioned by the authors in Ref.\,\cite{Mishra:2019pzq}, the presence of such localized bump leads to a large peak in the curvature power spectrum without significantly affect $n_s$ and $r$ on CMB scales, in contrast to polynomial USR models in which generating higher mass black holes severely affects $n_S$ and $r$.

In our case, the feature is modeled as a Gaussian bump,
\be
    V_{\rm bump}(\phi) 
    = V_0\, \delta\, \exp\left(- \frac{(\phi - \phi_0)^2}{2\sigma_\phi^2}\right)\,,
\ee
where $\delta$ characterizes the relative height of the feature, $\phi_0$ its location, and $\sigma_\phi$ its width. In this work, we focus on the case $\delta > 0$ (i.e., bumps), although features with $\delta < 0$ (dips) may also lead to spectral enhancement.

In the absence of the feature ($\delta = 0$), the field evolution for the logarithmic potential \eqref{eq:V_log+bump} can be solved analytically in the SR approximation. Moreover, this solution can be accommodated in the presence of the feature by implementing a simple shift in the number $e$-folds, as we'll show next.

\begin{figure}[!t]
	\centering
\includegraphics[width=0.495\textwidth]{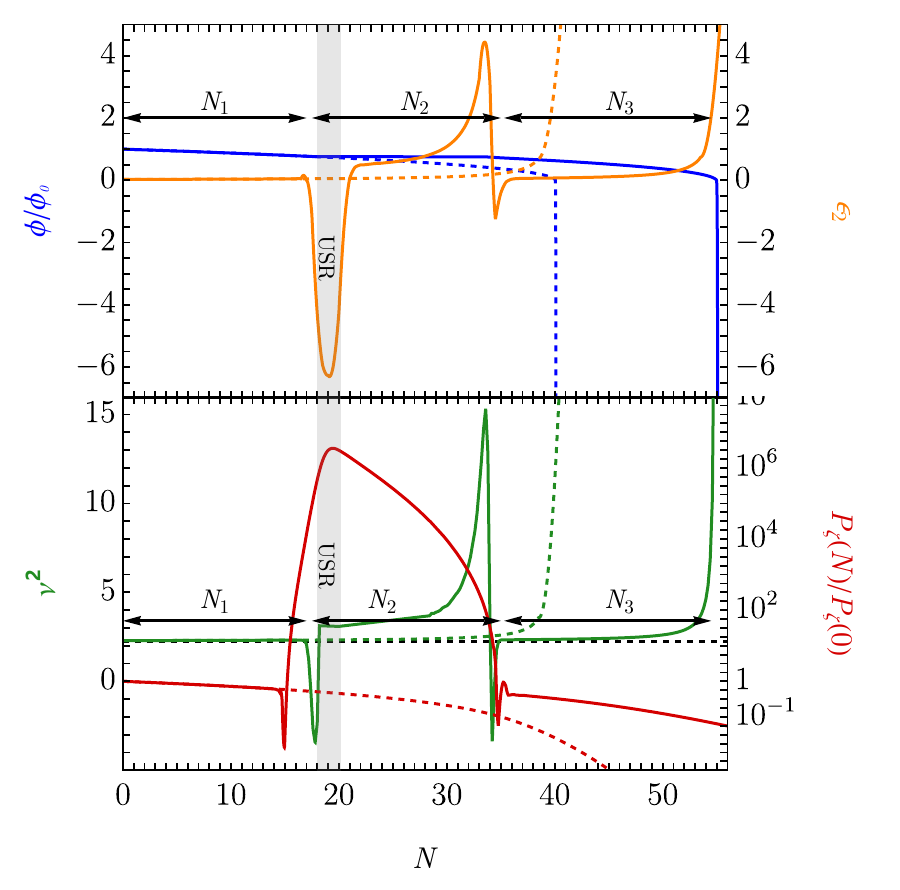}
\vspace{-5mm}
	\caption{ \justifying \it
Evolution of the parameter $\nu^2$ (green lines), the normalized curvature power spectrum $P_\zeta(N)/P_\zeta(0)$ (red lines), the field evolution (blue lines), and the second Hubble parameter $\eps_2$ (orange lines) as functions of the number of $e$-folds $N$. Solid lines correspond to the model including the localized bump, while dashed lines refer to the smooth potential without the bump. The two SR phases are labeled as $N_1$ and $N_3$, separated by a transient USR and CR stage whose beginning and end are given by the condition $\nu^2 \simeq 9/4$. The USR phase is indicated with a grey region.}
\label{fig:evo}
\end{figure}

By construction, the number of $e$-folds can be obtained from the first SR parameter
\be
\label{eq:N_SR}
    N
    = \int \frac{\td\phi}{\sqrt{2\eps_1}}\,.
\ee
To include the contribution of the bump, we must divide the range of integration into epochs during which the SR approximation ($\eps_1 \approx \eps_V$) can be applied, so the total number of $e$-folds, from the horizon exit of CMB scales to the end of inflation, is
\be
\label{eq:e-folds_toy}
    N_{\rm tot}
    \approx \underbrace{\int^{\phi_1}_{\phi_2} \td\phi\,\frac{V}{\pd_{\phi}V}}_{= N_1} 
    + \underbrace{\int^{\phi_2}_{\phi_3} \frac{\td\phi}{\sqrt{2\eps_1}}}_{= N_2}
    + \underbrace{\int^{\phi_3}_{\phi_{\rm end}} \td\phi\,\frac{V}{\pd_{\phi}V}}_{= N_3}\,,
\ee
where $\phi_1$ is the field value at the onset of inflation, while $\phi_2$, and $\phi_3$ correspond to immediately before and after the feature where the SR approximation breaks down. $\phi_{\rm end}$ marks the end of inflation. This is illustrated in the top panel of Fig.~\eqref{fig:evo}. As the feature is localized and the field excursion at the feature is small, $N_1$ and $N_3$ can be estimated purely from the logarithmic contribution of the potential~\eqref{eq:V_log+bump}. Moreover, inflationary observables at the CMB scales will match the predictions one would obtain with $N_{\rm SR} = N_1+N_3$ $e$-folds from the purely logarithmic potential, without the feature. This is the clearest demonstration of the approach outlined at the beginning of this section.
\begin{figure}[!t]
	\centering
\includegraphics[width=0.495\textwidth]{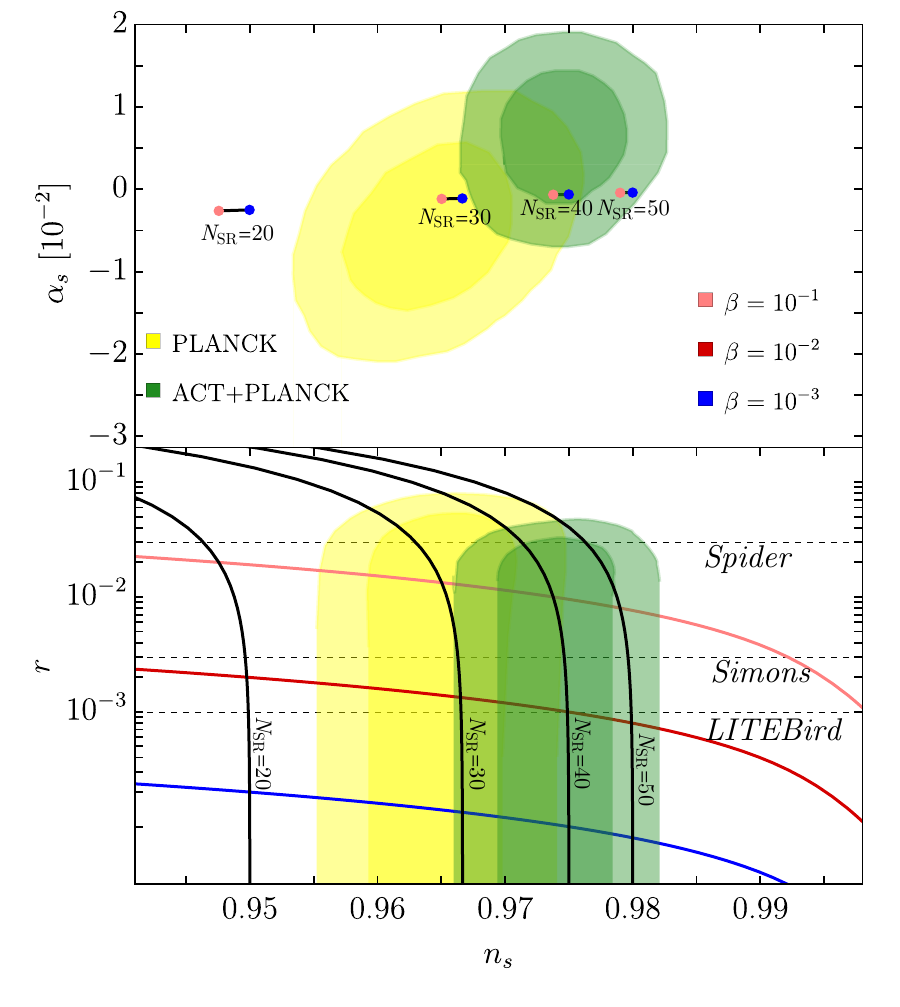}
\vspace{-7mm}
	\caption{ \justifying \it Constraints in the ($r,n_s$) plane (bottom panel) and ($\alpha_s,n_s$) plane (top panel) using the analysis of Planck data~\cite{Planck:2018vyg} (yellow regions) and the ACT+Planck~\cite{ACT:2025tim} (green regions). The plot includes 68\% and 95\% confidence regions. The colored lines show how the CMB observables change with varying $\beta$, while the black lines show the dependence on $N_{\rm SR}$. Prospective sensitivities on $r$ of experiments such as Spider~\cite{SPIDER:2017xxz}, Simons Observatory~\cite{SimonsObservatory:2018koc}, and LiteBIRD~\cite{Matsumura:2013aja} are indicated by the gray dashed lines.
 }
\label{fig:LogCMBConstraint}
\end{figure}

For the purely logarithmic part of the potential \eqref{eq:V_log+bump}, the integral \eqref{eq:N_SR} can be evaluated in SR, and inverting it gives (see App.\ref{app:toyexpl})
\be
\label{eq:phi(N)_toy}
    \phi(N) 
    = \frac{2\sqrt{N+N_0}}{\sqrt{W\left[4(N+N_0)\exp\left(2/\beta - 1 \right)\right]}}\,,
\ee
where $W$ denotes the Lambert $W$ function and $N_0$ is a constant. This allows us to estimate the inflationary observables, which, in the limit $\beta \to 0$, are well approximated by\footnote{As shown by the last expression, unlike the Starobinsky model \cite{Starobinsky:1980te}, where $n_s \simeq 1 - 2/N$ is too small compared to recent data, this model yields to $n_s\simeq1 - 1/N$, bringing it into agreement with the ACT results~\cite{ACT:2025fju,ACT:2025tim} for $N\in[50,55]$. The same behavior can be obtained by interpreting the logarithmic potential as the $\alpha \to 0$ limit of a polynomial potential, i.e $V(\phi) = \lambda_\alpha \phi^\alpha$, in the Palatini formulation \cite{Dioguardi:2025vci}.
}
\be\label{eq:app1}
    r = \frac{4\beta}{N}, \qquad
    n_s = 1 - \frac{1}{N}, \qquad
    \alpha_s = -\frac{1}{N^2}.
\ee

As discussed earlier, the SR evolution can be split into two phases, occurring before and after an USR or CR phase. These SR phases are delimited by the condition $\nu^2 \simeq 9/4$ so that $n_s \approx 1$. The spectral index of modes existing during the CR phase, that is, at scales following the peak, can be estimated from Eq.\eqref{eq:pot_SR_params}
\be\label{eq:ns_CR_toy}
    n_{s,\rm CR} 
    \approx 1 + 2\eta_{V}(\phi_0) 
%   = 1 - \frac{\delta}{\sigma_\phi^2(1 + \beta \ln\left(\phi_0/\MPl\right))}
    \approx 1 - \frac{\delta}{\sigma_\phi^2}\,,
\ee
where we assumed a small $\beta$ and $\eps_1$. Thus, the slope, and thereby also the length of the constant-$\nu$ phase, can be controlled by the parameters of the feature. %In particular, flatter spectra require $\delta \ll \sigma_\phi^2$.

Fig.~\ref{fig:evo} illustrates the evolution of the parameter $\nu^2$ (green) and the normalized scalar power spectrum (red), with and without the bump (solid and dashed lines, respectively). During the SR stages (denoted as $N_1$ and $N_3$), one finds $\nu \simeq 3/2$, with significant deviations occurring only near the bump or the end of inflation. Nevertheless, one can observe that $\nu$ remains nearly constant during the crossover from USR to CR, as illustrated in Fig.~\ref{fig:evo}. Moreover, notice that in Fig.~\ref{fig:evo}, when crossing from CR back to the final SR phase, $\nu^2$ has a sharp jump, which results in spectral oscillations at the beginning of the SR phase.

\begin{figure}[!t]
	\centering
\includegraphics[width=0.49\textwidth]{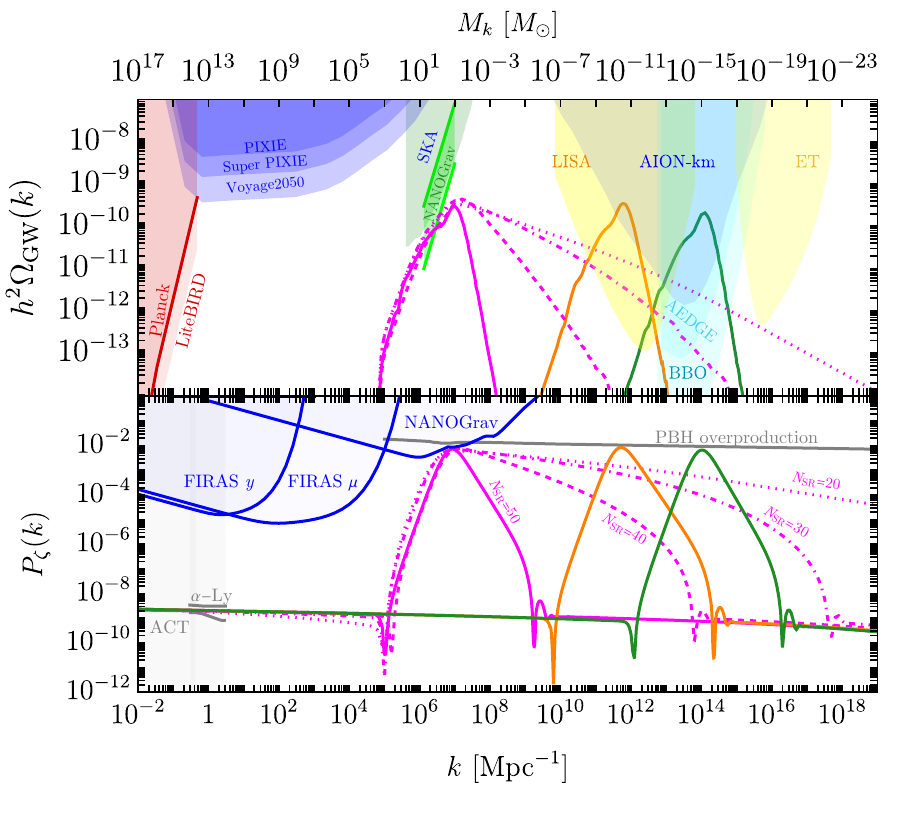}
	\caption{ \justifying \it Summary of the inflationary predictions and associated observational consequences for the toy model using the benchmark value $\beta=0.05$.
    \textbf{Bottom panel:} The curvature power spectra for  configurations with fixed $N_{\rm tot} \simeq 55$ and amplitude of the main peak $\mathcal{P}_\zeta(k_{\rm pk})\simeq10^{-2}$, but with different $N_{\rm SR}$ ( in magenta with different dashing) and with different $k_{\rm pk}$ (different colors). \textbf{Top panel:} The resulting SIGW spectra for the power spectrum shown in the bottom panel are color-coded accordingly.
}
\label{fig:Toy}
\end{figure}

Some example spectra obtainable from the potential~\eqref{eq:phi(N)_toy} are shown in Fig.~\ref{fig:Toy} together with their SIGW spectra. The configurations were constructed by tuning $N_{\rm SR}$ so that the inflationary observables fall within current observational bounds, while ensuring that the intermediate stage is long enough to satisfy $N_{\rm tot} \in [50, 55]$. The lower bound ensures successful inflation, while the upper bound\footnote{
The upper bound can be relaxed in scenarios featuring a prolonged reheating phase with an effective equation-of-state parameter $w>1/3$, while it becomes more stringent for scenarios with $w<1/3$~\cite{Allegrini:2024ooy}.
Although in this work we assume a standard thermal history with instantaneous reheating, it is instructive to note that, if we require $N_{ 2} \sim 3$ and $N_{\rm tot} \gtrsim 55$, it implies $N_{\rm SR} \gtrsim 52$, which would shift the inflationary observables, such as $(n_s,\alpha_s)$, outside the ACT constraints. Moreover, we stress that a prolonged reheating phase can generically affect the pre-recombination history of the Universe, thereby affecting both CMB observables~\cite{Munoz:2014eqa,Drees:2025ngb,Zharov:2025evb,Liu:2025qca} and the interpretation of PTA data in terms of scalar-induced gravitational waves~\cite{Liu:2023pau,Liu:2023hpw}. 
} 
avoids issues related to scale re-entry in standard reheating scenarios~\cite{Allegrini:2024ooy}. In all of these numerical examples, we find that the SR estimates predict the inflationary observables to an excellent accuracy, that is, the CMB observables can be estimated using Eq.~\eqref{eq:app1} (see also Eq.~\eqref{eq:eps_i_toy}) where $N$ is replaced by $N_{\rm SR}$.

Fig.~\ref{fig:LogCMBConstraint} shows the evolution of $(n_s, r, \alpha_s)$ for different benchmark values of $\beta$ and $N_{\rm SR}$ using the SR estimates in Eq.~\ref{eq:app1}. We find that $N_{\rm SR} \gtrsim 30$ is sufficient to match the ACT+Planck dataset at the $2\sigma$ level when $\beta < 0.1$. This implies that the potential~\eqref{eq:phi(N)_toy} can explain asteroid mass PBHs consistent with the ACT+Planck dataset even when $N_{3} = 0$, while solar-mass PBHs ($N_1 \approx 17$) would require another phase of SR inflation after the USR/CR phase ($N_{3} \gtrsim 13$).

Let us take a closer look at the benchmark models with $N_{\rm SR} \in \{20, 30, 40, 50\}$ shown in Fig.~\ref{fig:Toy} by the purple lines. To obtain these scenarios, the Gaussian bump was inserted in a way that guarantees $N_{\rm tot} \simeq 55$ and a power spectrum peak reaching $\mathcal{P}_\zeta (k_{\rm pk}) \simeq 10^{-2}$ at some scale $k_{\rm pk} \approx 10^{7} {\rm Mpc}^{-1}$. One can observe that the duration of the USR+CR phase $N_2$ can be tuned using the spectral index $n_{s,\rm CR}$ during CR -- a steeper slope yields a lower $N_2$ and thus a longer $N_{\rm SR}$. In this model, $n_{s,\rm CR}$ was given by Eq.~\eqref{eq:ns_CR_toy}.

The top panel of Fig.~\ref{fig:Toy} displays the SIGW spectra associated with each configuration. Requiring configurations compatible with the current ACT observations (solid and dashed lines), i.e. $N_{\rm SR} \simeq 40-50$, produces narrow-width power spectra and consequently a SIGW signal that do not span over a large range of frequencies. Nevertheless, configurations that produce asteroid-mass PBHs ($M_k \sim 10^{-15} M_\odot$) may be tested simultaneously by future observatories such as LISA~\cite{LISA:2022kgy}, BBO~\cite{Yagi:2011wg}, AION-km~\cite{Badurina:2019hst,Badurina:2021rgt,Abdalla:2024sst} and AEDGE~\cite{AEDGE:2019nxb,Badurina:2021rgt}.

The configuration with a relatively small value of $n_s \lesssim 0.965$ (dot-dashed line), i.e., $N_{\rm SR} \simeq 30$, corresponds to a broad-band SIGW signal spanning frequencies from the nHz regime, which is observed by current experiments (e.g., NANOGrav~\cite{NANOGrav:2023gor}) and future facilities (e.g., SKA~\cite{Zhao:2013bba,Babak:2024yhu}), up to the mHz-Hz regime, testable by upcoming detectors such as LISA~\cite{LISA:2022kgy}, BBO~\cite{Yagi:2011wg}, AION-km~\cite{Badurina:2019hst,Badurina:2021rgt,Abdalla:2024sst}, AEDGE~\cite{AEDGE:2019nxb,Badurina:2021rgt}, and ET~\cite{Abac:2025saz}. While remaining compatible with Planck constraints, such multi-band signals can also arise in other USR scenarios, such as polynomial inflation~\cite{Frosina:2023nxu,Allegrini:2024ooy} discussed in the section below.

For completeness, the top panel of Fig.~\ref{fig:Toy} shows also the sensitivity curves of present or future experiments seeking ultra-low frequency GWs, that may, for instance, arise from the ringdown phase of very heavy PBHs~\cite{DeLuca:2025uov,Yuan:2025bdp}. These include sensitivity from future proposed CMB experiments looking for B-modes, such as PIXIE~\cite{Kogut:2011xw}, Super Pixie~\cite{Kogut:2019vqh} and Voyage2050~\cite{Chluba:2019nxa}; current excluded region (Planck~\cite{Planck:2018jri,Tristram:2020wbi,BICEP:2021xfz}) and future sensitivity (LiteBIRD~\cite{Hazumi:2019lys}) from space-based experiments targeting the indirect detection of primordial gravitational waves through the measurement of the CMB B-modes.

\subsection{Non-minimal polynomial inflation}

In the following, we will consider a canonical ($K(\phi)= 1$) non-minimally coupled Jordan frame scalar with an effective potential, so that\footnote{A similar model has been considered in Ref.~\cite{Catinari:2025itc}, where it was argued that combining it with a non-minimally coupled Higgs field can lead to problems with naturalness. This potential issue is outside the scope of this paper.}
\bea
\label{eq:potential}
    \Omega(\phi) 
    &= 1 + \sum_{n\geq 1} \xi_n \left( \frac{\phi}{\MPl} \right)^n\,,
    \\
    V(\phi) 
    &= \MPl^4 \sum_{n\geq 1} a_n \left( \frac{\phi}{\MPl} \right)^n\,,
\eea
where $\xi_n$ are non-minimal couplings with gravity and $a_n$ are dimensionless coefficients. We can assume $a_1 = 0$ without loss of generality, as an appropriate shift in the field can always absorb this linear term. Although the non-linear term has been dropped in earlier studies~\cite{Kannike:2017bxn, Ballesteros:2020qam, Frosina:2023nxu}, we will include it for generality. In particular, linear terms are expected to be generated by quantum corrections as their absence is not enforced by any symmetry, and thus omitting such terms is generally not justified.  However, we do not consider non-minimal terms of dimension greater than four although they may be present in an effective field theory setting, considered here. To achieve a sufficiently high scalar spectral index, it was argued that the potential should contain terms of at least up to dimension 6~\cite{Ballesteros:2020qam}. In all, the model contains seven independent dimensionless parameters $\xi \equiv \xi_2$, $\xi_1$ and $a_i$, $i \in \{2,3,4,5,6\}$ of which one can be absorbed into the overall scaling of the potential. We choose this to be the coefficient of the quadratic coupling $a_4$.

Introducing the dimensionless Jordan frame field $x \equiv \phi / \MPl$, the Einstein frame potential reads
\bea
\label{eq:Pot1}
    \bar V(x) 
    \!=\!\frac{a_4 \MPl^4}{\left(1\!+ \xi_1 x +\xi x^2\right)^2}
    \left(
    \tilde{a}_2 x^2\!+\!
    \tilde{a}_3 x^3\!+\!
    x^4\!+\!
    \tilde{a}_5 x^5\!+\! 
    \tilde{a}_6 x^6
    \right)\,,
\eea
where we defined $\tilde{a}_k \equiv a_k/a_4$ to absorb the quartic coupling into the prefactor.  The coefficient~\eqref{eq:barV,barK} of the non-canonical kinetic term is 
\be
\bar K = \frac{1+\xi_1+ \xi x^2 + \frac{3}{2}(\xi_1 + 2 \xi x)^2}{(1+ \xi_1 x+\xi x^2)^2}.
\ee

We will separately look at non-minimally coupled models ({\bf NMC}) and minimally coupled ({\bf MC}) models, which correspond to the $\xi_i=0$ special case of the NMC models. The production of PBHs consistent with a viable CMB spectrum in the case of a minimally coupled scalar field can be achieved by imposing a second inflection point~\cite{Allegrini:2024ooy}. In such scenarios, the first inflection point flattens the potential, ensuring compatibility with large-scale CMB observations, while the second generates a peak in the power spectrum required for PBH formation.

\begin{figure*}[t!]
	\centering
\includegraphics[width=0.85\textwidth]{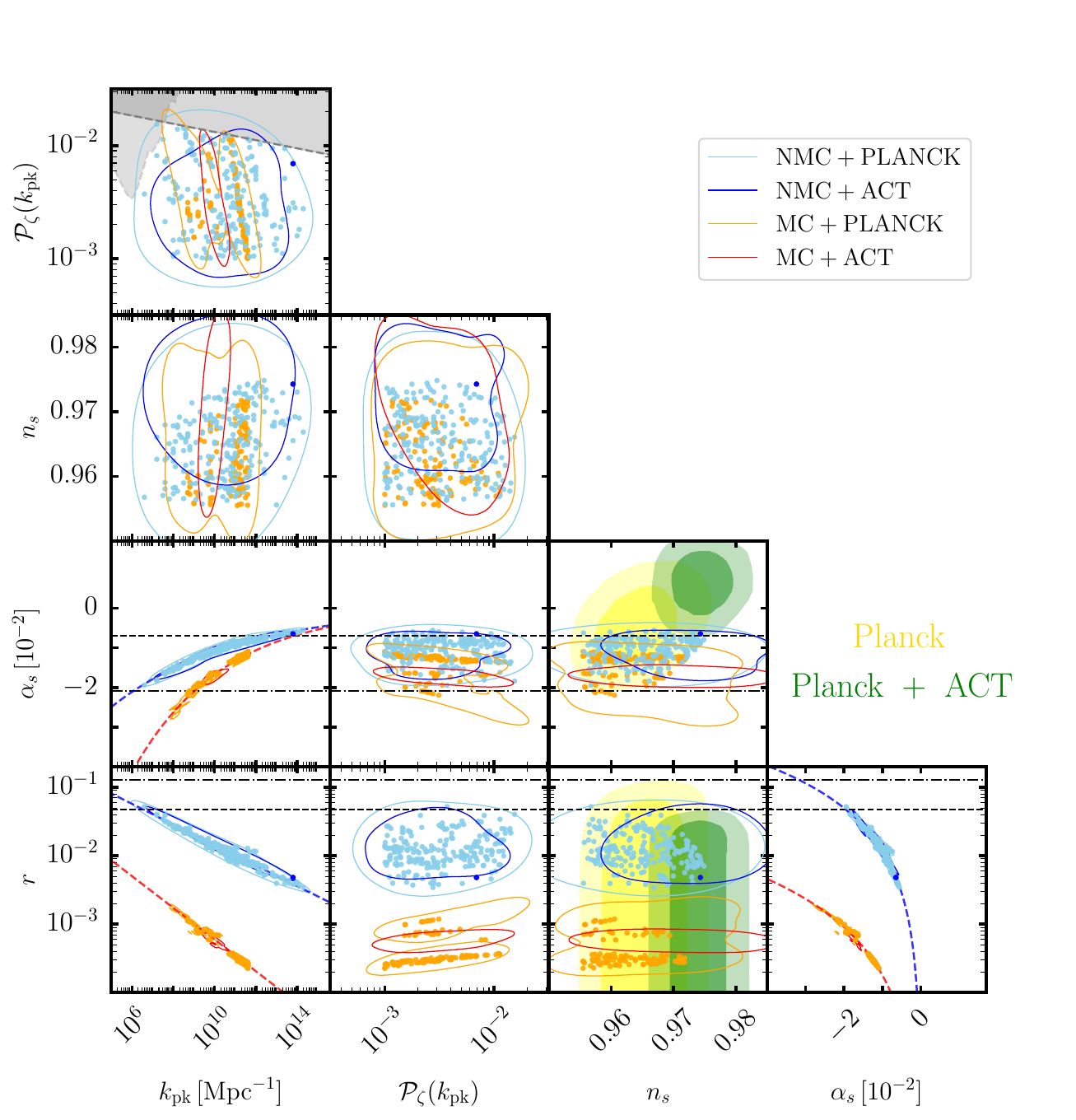}
	\caption{ \justifying  \it Output parameters from the \textbf{NMC} (blue) and \textbf{MC} (red) models from the MCMC scan. The dots represent model configurations that fall within the 68\% or 95\% confidence regions from Planck data~\cite{Planck:2018vyg} (yellow) and ACT+Planck~\cite{ACT:2025tim} (green). The corresponding $2\sigma$ bounds are indicated by black dashed lines. Configurations are further required to exhibit a power spectrum peak compatible with the NANOGrav and PBH overproduction constraints (grey regions in the top-left panel). Vivid colors indicate the $2\sigma$ contours for the configurations obtained from the Planck+ACT MCMC, while faded colors indicate those obtained from the Planck MCMC (see App.~\ref{app1} for further details).
 }
\label{fig:Corner}
\end{figure*}

%-------------------------------------------------------------------------------
\section{Confronting PBHs with CMB constraints}
\label{sec:fit}
%-------------------------------------------------------------------------------

\begin{figure*}[t!]
\centering
\begin{minipage}{.49\textwidth}
  \centering
  \includegraphics[width=\linewidth]{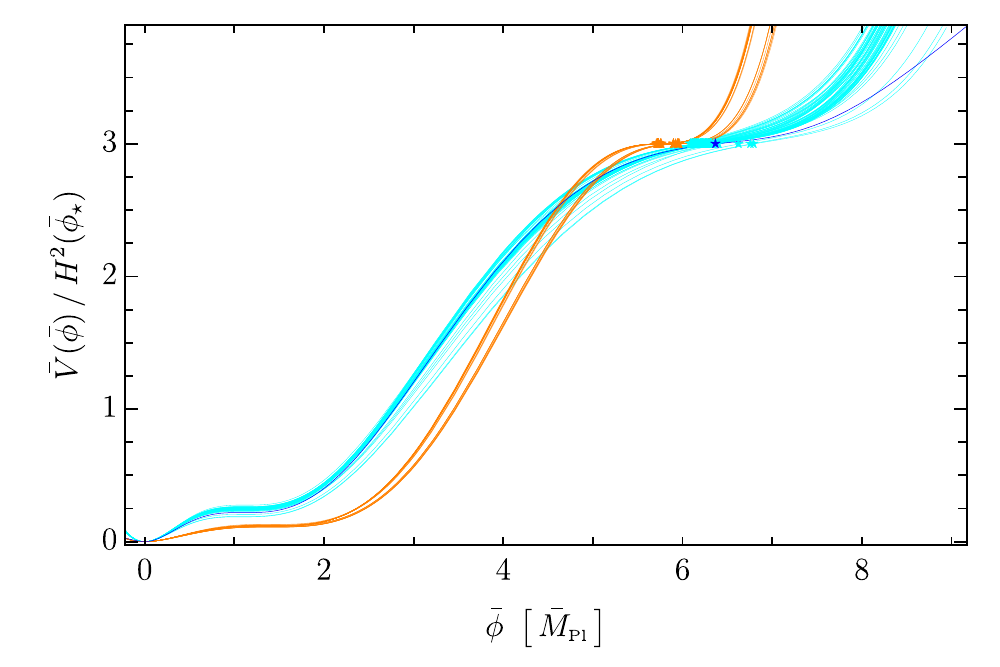}
\end{minipage}\hfill
\begin{minipage}{.51\textwidth} 
  \centering
  \includegraphics[width=\linewidth]{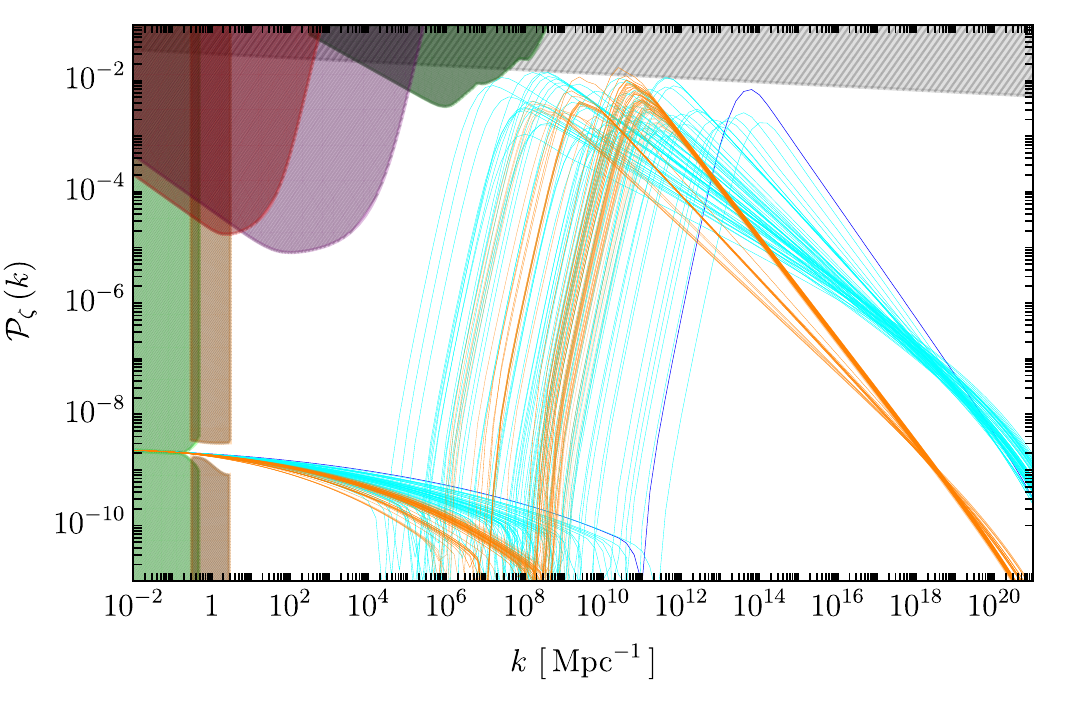} 
\end{minipage}
\caption{\justifying \it Potentials (left) and curvature power spectra (right) corresponding to the model configurations highlighted as dots in Fig.~\ref{fig:Corner}, shown using the same color scheme and rescaling the opacity of the lines to match the likelihood. For clarity, we present only 70 benchmark configurations from the MC+Planck and NMC+ACT cases. In the left panel, a small star marks the value of $\phi(N_*)$ at the reference scale for each potential. The constraints displayed in the right panel are the same as those shown in Fig.~\ref{eq:PS}.}
    \label{fig:PS+POT}
\end{figure*}

Now we move to the more realistic polynomial inflation introduced in Section~\ref{sec:modelbulding}. We perform a comprehensive scan over the parameter space with the help of an MCMC algorithm, to uncover models for a broad range of PBH masses. MCMC scans are performed with the \texttt{emcee}~\cite{Foreman-Mackey:2012any} ensemble sampler. These correspond to characteristic scales from solar-mass PBHs at $k_{\rm pk} \simeq 10^{5}\,{\rm Mpc}^{-1}$, to asteroid-mass PBHs at $k_{\rm pk} \simeq 10^{15}\,{\rm Mpc}^{-1}$. The reported inflationary observables $(n_s,\,\alpha_s,\, r)$ as well as the position $k_{\rm pk}$ and the amplitude $P_{\zeta}(k_{\rm pk})$ of the PBH-generating peak are obtained from the Mukhanov-Sasaki equation~\eqref{eq:MukSas}.
More details on the MCMC analysis are reported in App.~\ref{app1}. 

The results are summarized in Fig.~\ref{fig:Corner}. The colored lines surround the region containing $95\%$ of the models found and thus illustrate the range of models obtained by the scan. The points correspond to individual models for which $(n_s,\,\alpha_s,\, r)$ lie within the $2 \sigma$ region of Planck (Cyan and Orange) or Planck+ACT (Blue and Red). The following points can be made:
\begin{itemize}
    \item[{\it i)}] The $k_{\rm pk} - P_{\zeta}(k_{\rm pk})$ panel maps out the most relevant range for GW experiments as the SIGW is directly related to the height of the peak. It shows the regions excluded by PBH overproduction and PTA observations (gray). We find that, omitting ACT data, $P_{\zeta}$ peaks can be generated over a wide range of scales (or frequencies), while the inclusion of ACT data leaves us with a single NMC model blue dot in the asteroid mass range.
    \item[{\it ii)}] The spectral index $n_s$ can vary across the entire observationally allowed range and does not exhibit significant correlations with other parameters. This behavior is evident from the second row and the third column. Therefore, $n_s$ is not a strong limiting factor for model selection, at least for the class of polynomial models considered in this scan.
    \item[{\it iii)}] $\alpha_s$ is negative and decreases predictably with decreasing $k_{\rm pk}$. This is one the main limiting factor for model section and thus of the key takeaways. A negative running can be reconciled with Planck data, but the Planck+ACT measurement excludes all but one model (blue dot) at the $2\sigma$ CL as noted above.

    As a result, we find that MC polynomial inflation for solar mass PBHs is incompatible with both the Planck and Planck+ACT measurements.
    
    \item[{\it iv)}] NMC models tend to yield a larger $r$ and $\alpha_s$ than the MC ones as can be seen on the bottom row.
    Both remain within $2\sigma$ of Planck+ACT in the $(n_s,r)$ plane, with MC inside $1\sigma$ throughout, while NMC is excluded for solar mass or heavier PBHs ($k_{\rm pk} \lesssim 10^6\,{\rm Mpc}^{-1}$) due to excessive $r$. The MCMC reveals a specific trade-off: a non-minimal coupling can increase $r$ in order to attain a smaller $\alpha_s$.
    
    \item[{\it iv)}] We find strong correlations between  $k_{\rm pk}$, $r$ and $\alpha_s$, and their relations depends on $\xi$. Since the MC models represent a special case of the NMC ones, this suggests that the observed correlations likely arise from the way the MCMC algorithm explores the parameter space, which must require fine-tuning in any USR model. Although the algorithm can identify models covering a wide range of phenomenological parameters, its ability to move efficiently through the parameter space is limited. In particular, since the NMC models found correspond to $\xi \in [0.25, 0.3]$, the algorithm is unable to reach the MC limit ($\xi = 0$) when starting from an NMC configuration. We will discuss these correlations in more detail in subsection~\ref{eq:correlations}.
\end{itemize}

For all models highlighted as dots in Fig.~\ref{fig:Corner}, the Einstein frame potentials (left) and curvature power spectra (right) are shown in Fig.~\ref{fig:PS+POT} where we used the same coloring scheme. An inflection point can be seen on the PBH scales as well as on the CMB scales. The latter is indicated by a star for each potential. Note that the potentials are plotted in terms of the Einstein frame field, and thus the ``stretching'' of the potential due to the field redefinition is accounted for this figure. The inflection point tends to flatten the Einstein frame potential between the two inflection points. As a result, the NMC model potentials have milder features at the CMB scales and lead to a lower $\alpha_s$.

Despite the similarity of the potentials, the power spectra in Fig.~\ref{fig:Corner} display a wide array of different peaks, as indicated by Fig.~\ref{fig:Corner}. This is not surprising because it is expected that the features of these peaks are determined by the local features of the potential around the lower inflection point. As the peak shape and position are sensitive to changes in these features, their variations are generally not resolved when plotting the global shape of the potential.

%-------------------------------------------------------------------------------
\subsection{Understanding correlations between $k_\textrm{pk}$ and the inflationary parameters}
\label{eq:correlations}

Let us take a closer look at the correlations between $k_{\rm pk}$, $r$, and $\alpha_s$ shown in Fig.~\ref{fig:Corner}. The immediate question is whether this correlation arises due to the variation in the duration of the first phase of SR, i.e., $N_1$, which is directly related to $k_{\rm pk}$ by Eq.~\eqref{eq:kN1} and for which $r$ and $\alpha_s$ are expected to have a relatively simple relation within a \emph{fixed} model as we saw, for example, in the toy model~\eqref{eq:app1}. However, this explanation may fail because Fig.~\ref{fig:Corner} depicts correlations between different models, and, indeed, as we shall show, this explanation turns out to be incorrect.

A quantitative relation of this correlation can be obtained by fitting the $k_{\rm pk}- r$ and $k_{\rm pk}-\alpha_s$ relations in Fig.~\ref{fig:Corner} for the MC and NMC scenarios using the form
\bea\label{eq:fit}
    r \simeq c_1\, e^{c_2\,N_1}\,, 
    \quad
    \alpha_s \simeq c_3 \, e^{c_4\, N_1}\,, \\
\eea
where $k_{\rm pk}$ is expressed using the duration of the first SR phase, $N_1$, using Eq.~\eqref{eq:kN1}. The numerical coefficients are $c_i = \{0.164, -0.214, -0.182, -0.091\}$ for the MC models and $c_i = \{0.601, -0.146, -0.065, -0.070\}$ for the NMC models. This implies that $r \propto -\alpha_s^{c_2/c_4}$, which yields exponents $c_2/c_4 =$ 2.4 and 2.1 for the MC and NMC models, respectively. In other words, $r$ and $\alpha_s$ have approximately a quadratic dependence. The corresponding trend lines are shown in the $k_{\rm pk}- r$, $k_{\rm pk}-\alpha_s$ and $r-\alpha_s$ panels of Fig.~\ref{fig:Corner} by a dashed line. The fitting functions~\eqref{eq:fit} capture the correlations very accurately.

\begin{figure}[!t]
	\centering
\includegraphics[width=0.48\textwidth]{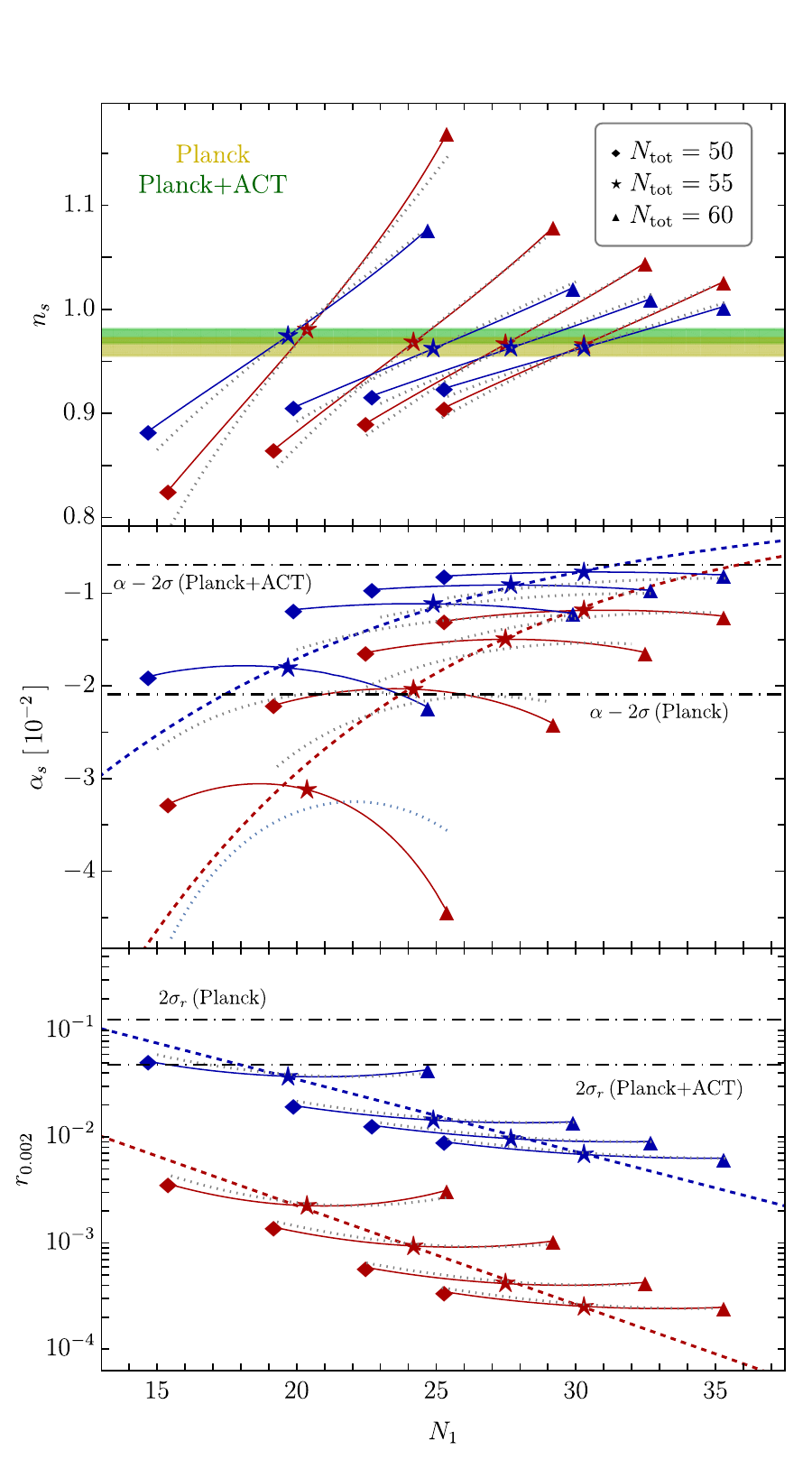}
	\caption{ \justifying \it  Comparison of the evolution of the inflationary parameters, $n_s$, $\alpha_s$, and $r$ for different positions of $N_1$ in the MC (red) and NMC (blue) models. We compare the behavior of some benchmark realizations of the two models obtained from the MCMC scan (solid lines) with the fits obtained in the corner plot of Fig.~\ref{fig:Corner} (dashed lines), where we expressed $N_1$ in terms of $k_{\rm pk}$. In each panel we highlight 95\% confidence levels from Planck data~\cite{Planck:2018vyg} and ACT+Planck~\cite{ACT:2025tim}, as well as inflationary parameters at fixed $N_\textrm{tot} = 50,\, 55, \, 60$.}  
\label{fig:nsalphar}
\end{figure}

A more detailed view of these correlations is provided by Fig.~\ref{fig:nsalphar}. It shows a few benchmark points from the set of models found by the MCMC scan (indicated by a star), in which the total number of $e$-folds was fixed to 55, but $N_1$ was allowed to vary. The solid lines show the change in inflationary observables when $N_{\rm tot}$ is allowed to vary from 50 to 60 $e$-folds. This is achieved by shifting the start of inflation, and thus the duration $N_2$ of the USR/CR phase of inflation is unaffected and only $N_1 = N_{\rm tot} - N_2$ changes. 

On the other hand, the colored dashed trend lines from Eqs.~\eqref{eq:fit} approximate the correlations found in the MCMC. The same trend lines are also shown in~Fig.~\ref{fig:Corner}. Fig.~\ref{fig:nsalphar}, therefore, shows that the variation in $N_1$ cannot explain the correlations observed in~\ref{fig:Corner}. It further follows that the inflationary models corresponding to the points in Fig.~\ref{fig:Corner} exhibit a quantitatively different time evolution, and thus the MCMC algorithm is capable of finding new models instead of simply varying the duration of the initial SR phase of the same model due to small changes in the properties of the peak.

Moreover, this analysis reveals the reason why $n_s$ appears to be uncorrelated and can be freely adjusted and why $\alpha_s$ and $r$ exhibit a strong correlation with $k_{\rm pk}$. That is, as shown in the top panel of Fig.~\ref{fig:nsalphar}, throughout the entire range of $N_1$ spanned by the MCMC, the value of $n_s$ can be adjusted with only a small change in the duration of the SR phase (and also $N_{\rm tot}$) while $\alpha_s$ and $r$ are relatively unaffected by it.

Comparing individual realizations of polynomial inflation (solid lines) at different $N_1$ with the general trends extracted from the fits in Fig.~\ref{fig:Corner} (dashed lines), one might be tempted to think that by adjusting the duration of inflation $N_\textrm{tot}$, it is possible to simultaneously increase $\alpha_s$ and decrease $r$ at a given $N_1$, thus improving agreement with the CMB constraints. However, as is clear from the top panel, such a tuning would quickly drive $n_s$ to values that are entirely incompatible with current constraints. Therefore, the top panel enforces that, to remain consistent with the allowed range of $n_s$, $N_\textrm{tot}$ can only be varied within approximately one e-fold. Consequently, the resulting values of $\alpha_s$ and $r$ will still be very close to the dashed-line predictions.

%-------------------------------------------------------------------------------
\subsection{CMB at an inflection point: why is $\alpha_s$  negative?}
\label{eq:inflection}

So far, we have looked at the behavior of $r(N_1)$ and $\alpha_s(N_1)$ using numerical solutions in MC and NMC models. To develop a quantitative analytic understanding, we can use the fact that all polynomial models exhibit a non-stationary inflection point close to CMB scales. This means that at CMB scales, inflation is expected to mimic standard inflection point inflation~\cite{Catinari:2025itc, Bernal:2021qrl, Drees:2021wgd, Drees:2025ane}. To make this connection more rigorous, let us consider the fourth-order Taylor expansion of the Einstein frame potential around the second inflection point at $\bar \phi = \bar \phi_{\rm i}$, that is, $\pd^2_{\bar \phi}\bar V (\bar \phi_{\rm i}) = 0$. We parametrize the potential as
\bea
\label{eq:quasi_inflection_point}
    \bar V(\delta) 
    &= \mu \left( \bar{V}_0 + \beta \delta + \delta^3 \right) + \mathcal{O}(\delta^4)\,,
\eea
where $\delta \equiv (\bar \phi-\bar \phi_{\rm i})/\MPl$, $\beta$ quantifies the deviation from a stationary inflection point, $\bar{V}_{0} \equiv \bar{V}(\bar \phi_{\rm i})/\mu$ characterizes the height of the potential at the inflection point and $\mu$ is a constant of proportionality.

The inflationary observables implied by the expansion \eqref{eq:quasi_inflection_point} at the second order in SR (for details, see App.\,\ref{app:IPI})
\bea\label{eq:SRexpansion}
    n_s &\simeq 1 + \frac{12 \delta_\star}{V_0} - \frac{(3\beta +12 C - 4)\beta}{V_0^2} \,, \\
    \alpha_s &\simeq  - \frac{12  \beta}{V_0^2} - \frac{36 \delta_\star^2}{V_0^2} \,, \\
    r &\simeq \frac{8 \beta^2}{V_0^2} + \frac{48 \delta_\star^2 \beta}{V_0^2} \,,
\eea
with the number of e-folds given by
\bea\label{eq:N1_IP}
    \delta_\star(N_1) 
    \simeq \frac{\sqrt{\beta}}{\sqrt{3}}\tan\left[\frac{\sqrt{3 \beta}}{V_0}(N_1 - N_0)\right]\,,
    %\\
    %&\simeq \frac{\beta}{V_0} (N_1-N_0) + \mathcal{O}(N_1-N_0)^2\,,
\eea
where $N_0$ is the number of $e$-folds it takes to roll from the first inflection point to the end of the first SR phase and $\delta_\star$ denotes $\delta$ at the horizon exit of the pivot scale $k_\star$. The constant $C \simeq-0.73$ arises in the second-order SR expansions and thereby exemplifies why the first-order approximation is not sufficient. These estimates were obtained assuming $\beta \ll V_0$ and $|\delta_\star| \ll 1$. Tab.\,\ref{tab:Parameters} shows that these conditions are satisfied for both the MC and the NMC models considered here.

\begin{table}[htp]
	\begin{center}%\vspace{-0.25cm}
    \def\arraystretch{1.2}
	\begin{adjustbox}{max width=.5\textwidth}
	\begin{tabular}{c>{\centering\arraybackslash}p{0.26\linewidth}>{\centering\arraybackslash}p{0.26\linewidth}>{\centering\arraybackslash}p{0.3\linewidth}}
      \hline
      &  $V_0$ & $\beta$ & $\delta_\star$ 
      \\
      \hline
      \bf{MC} & 
      $[5.5, \, 6.1]$
      & 
      $[0.03, \, 0.1]$
      &
      $[-0.028,\, -0.021]$
      \\
      \bf{NMC} & 
      $[39.6, \, 40.1]$
      & 
      $[1.1, \, 2.7]$ 
      &
      $[-0.156,\, -0.152]$
      \\
      \hline
    \end{tabular}
	\end{adjustbox}
	\end{center}\vspace{-0.0cm}
    \caption{\it 
Parameter ranges the benchmark potentials shown in Fig.~\ref{fig:nsalphar} in terms of the cubic expansion of Eq.\,\eqref{eq:quasi_inflection_point}.}\label{tab:Parameters}
\end{table} 
    
The analytic estimates~\eqref{eq:SRexpansion} are shown in Fig.~\ref{fig:nsalphar} by gray dashed lines. One can see that they reproduce the numerically obtained $r$ and $n_s$ very well and also that $\alpha_s$ is captured qualitatively. The inaccuracy in the $\alpha_s$ estimates is due to the omission of the quartic order in the expansion~\eqref{eq:quasi_inflection_point}. When $\delta_\star$ is large, the effect of such terms on $\alpha_s$ would  be larger than for $r$ and $n_s$. In fact, that is exactly what can be observed in Fig.~\ref{fig:nsalphar}: the numerical and analytical estimates of $\alpha_s$ are in good agreement for the values of $N_1$ where $n_s \approx 1$, that is, when we are close to the inflection point and start to disagree more when we deviate significantly from scale invariance. As a result, the estimates~\eqref{eq:SRexpansion} are expected to work well in nearly scale invariant scenarios that explain the observed the CMB spectrum.

The analytic estimates~\eqref{eq:SRexpansion} can also explain the approximate relation $r \propto -\alpha_s^2$ that we observed in the bottom right corner of Fig.~\eqref{fig:Corner}. For this, we must also consider Tab.\,\ref{tab:Parameters}, which shows that the MCMC scan varied mostly the parameter $\beta$ while $\delta_\star$ and $V_0$ remained relatively constant. Since $\alpha_s \propto \beta$ and $r \propto \beta^2$ at the leading order, we reproduce the numerical fit~\eqref{eq:fit}. The small deviations from an exact $r \propto -\alpha_s^2$ relation found in the fit~\eqref{eq:fit} can be attributed to changes in other variables. 

Thus, as we hypothesized in the previous section, the $r-\alpha_s$ correlations in Fig.~\eqref{fig:Corner} are a result of how the MCMC maps the parameter space of the models. The tuning at the lower inflection point inherent to USR models can restrict the ability of MCMC algorithms to cover the full parameter space. Therefore, a likely reason for why $\beta$ can vary the most, is that it has the smallest impact on small field physics when compared to, e.g., $V_0$ and is thus less affected by the tuning at the lower inflection point.

Finally, to obtain an even simpler quantitative description, it is sufficient to note that an inflection point implies that there is a point in the evolution of a slowly rolling field at which $\eps_2$ vanishes. Indeed, this was explicitly confirmed by the SR analysis (see also App.\,\ref{app:IPI}). Given that $\eps_2$ will continuously change its sign as it crosses the inflection point, we can expand
\be
    \eps_2(N)\approx -\alpha_{\star} (N\!-\!N_0) + \mathcal{O}(N\!-\!N_0)^2\,.
\ee
where $\alpha_{\star}$ is a constant. Noting the hierarchy among the Hubble parameters at the CMB scales $\eps_1 \ll \eps_2 \lesssim \eps_3$, we find that 
\be
    \alpha_s 
    \approx -\eps_3\eps_2 
    = -\pd_N \eps_2 
    = \alpha_{\star} + \mathcal{O}(N - N_0)\,,
\ee
that is, the constant $\alpha_{\star}$ describes the running of the spectral index close to the inflection point. Given that $\eps_2 \simeq -2\eta_V + 4\eps_1 \simeq -2\eta_V$ when $\eps_1 \ll \eps_2$, we obtain after expanding in $N-N_0$
\bea
    \eps_2 
    &\!\approx\! 
    -2\left. \pd_{\bar \phi}\eta_V \pd_N \bar\phi\right|_{N=N_0} (N\!-\!N_0) + \mathcal{O}(N\!-\!N_0)^2
    \\
    &\approx 2\xi^2_V(\bar \phi_0) (N\!-\!N_0) + \mathcal{O}(N\!-\!N_0)^2\,,
\eea
where we used $\pd_N \bar\phi \approx -V'/V$ at the leading order in SR and the third SR parameter is calculated at the inflection point $\bar \phi_0 \equiv \bar \phi(N_0)$ (i.e., $V''(\bar \phi_0)=0$). As a consistency check, for the expanded potential~\eqref{eq:quasi_inflection_point} we have
\be
    \xi^2_V(\bar \phi_0) = \frac{6\beta}{V_0^2} = -\frac{\alpha_\star}{2}\,,
\ee
where $\alpha_{\star}$ can be read off Eq.\eqref{eq:SRexpansion}. Since $V_0^2>0$, the sign of $\alpha_s$ is determined by $\beta$. The potential is asymmetric \eqref{eq:quasi_inflection_point}, so in the current setup, the field must start rolling at some $\bar\phi>0$ and then proceed towards $\phi=0$. When starting close to the inflection point, this is possible only when $\beta > 0$. Otherwise, the potential would have a local maximum at $\bar\phi < \bar\phi$, and the field would get stuck.

As a result, $\alpha_s < 0$ and $|\alpha_s|$ can be reduced by increasing $V_0$ or reducing $\beta$. Table~\ref{tab:Parameters} shows that both $V_0$ and $\beta$ are larger in the NMC models than in the MC models. However, the quadratic dependence $V_0$ prevails over the linear dependence $\beta$, which is why the NMC scenarios produce systematically lower $|\alpha_s|$ and are therefore in better agreement with CMB observations.

Finally, we stress that we find that $\alpha_s < 0$ for polynomial models and the logarithmic toy model~\eqref{eq:V_log+bump} (see Eq.~\eqref{eq:app1}). Therefore we do not claim, that $\alpha_s < 0$ is a universal prediction of inflationary models for CMB and PBHs. In fact, it is likely that sufficiently complex potentials (i.e, with more independent terms and free parameters) would allow more freedom in tuning the primordial curvature power spectrum and thus permit constructing models with arbitrary $r$, $n_s$ and $\alpha_s$.
Even so, the polynomial models considered here represent the most complex ones considered in the literature so far. Within this class, the MCMC procedure selects a subset of models characterized by a second, non-stationary inflection point at large field values, for which $\alpha_s < 0$ emerges as a robust outcome, largely independent of the detailed shape of the potential (e.g, the polynomial order). It remains an interesting open issue whether one can devise an alternative class of inflationary models of comparable complexity that could naturally evade a negative $\alpha_s$.

\vspace{-3mm}
\section{Conclusions}
\label{sec:conclusions}
%-------------------------------------------------------------------------------

In this paper, we have demonstrated that the running of the scalar spectral index plays a crucial role in determining the allowed mass distribution of PBHs within the framework of single-field inflation.
In this context, polynomial models for PBH production tend to favor configurations with a non-stationary inflection point close to the CMB scales, which implies a negative $\alpha_s$. Therefore, the recent ACT dataset, which prefers a positive $\alpha_s$, places tighter constraints on the viable PBH mass range.

In order to analyze this issue, we have provided a systematic framework for constructing single field inflationary models capable of generating a sizable population of PBHs while remaining consistent with the most recent CMB constraints. 

Firstly, in Sec.\,\ref{sec:modelbulding}, we outlined a practical and flexible recipe for model construction in which a transparent connection exists between the potential features, the dynamics of the inflaton, and the resulting curvature power spectrum. Starting from a potential that reproduces CMB observables within a reduced number of $N_1$ $e$-folds, one can introduce a localized feature to induce a temporary USR stage of duration $N_2$, followed by a final SR regime of $N_3$ $e$-folds. In this framework, the total number of SR $e$-folds, $N_{\mathrm{SR}}=N_1+N_3$, controls the inflationary observables $(n_s,\,r,\,\alpha_s)$, while $N_1$ sets the mass scale of PBH formation through the relation $k_{\mathrm{pk}}\!\propto\!e^{N_1}$.

By applying these approaches to different inflationary models, we have shown that the interplay between $N_{\mathrm{SR}}$ and $N_1$ determines whether a model can simultaneously reproduce the CMB spectrum and generate a significant PBH abundance. We found that although certain configurations of these two models yield values of the standard inflationary parameters $(n_s, r)$ that are consistent with current observational constraints, the running parameter $\alpha_s$ often falls outside the experimentally allowed range. This mismatch imposes stringent upper bounds on the maximal PBH mass in these scenarios.

Secondly,in Sec.\,\ref{sec:fit}, we utilized a Markov Chain Monte Carlo method for mapping the model space to scan the parameter space of non-minimally coupled polynomial inflation. This approach relies on constructing a mock-likelihood that accounts for CMB posteriors together with a condition imposing the production of PBHs. Using this method, we were able to find scenarios that covered a wide range of inflationary observables $(n_s,\,r,\,\alpha_s)$. However, since the condition required to produce a sizable PBH abundance introduces degeneracies in the parameter space, we were not able to identify models that differ significantly in their non-minimal couplings. Nevertheless, the adopted approach remains rather simplistic, yet it shows promise for being adaptable to a much broader class of USR models than those considered here, once the issue of degeneracy is properly addressed. Developing effective methods to overcome this limitation will be crucial for performing reliable statistical inference with potential future GW background data.

Focusing exclusively on the $\alpha$-attractor E model, a similar tension has been observed in the asteroid mass range~\cite{Frolovsky:2025iao}. Unlike the aforementioned work, our analysis generalizes the discussion in a model-independent way and explores the full range of PBH mass allowed by current observational constraints.
Our findings highlight that generating PBHs in the (sub)solar mass range, potentially observable by LIGO-Virgo-KAGRA~\cite{LIGOScientific:2018mvr, LIGOScientific:2020ibl, KAGRA:2021vkt,Pujolas:2021yaw,LIGOScientific:2025slb} and Einstein Telescope~\cite{Branchesi:2023mws, Franciolini:2023opt,Abac:2025saz} collaborations, within single-field inflationary models remains a challenging task.

Several extensions of our work are possible.
First, it is important to identify USR models in which is possible to produce a positive running $\alpha_s$.  In addition, we note that a population of sub-solar mass PBHs produced in USR models inevitably gives rise to a SIGW signal in the nHz frequency range. This signal could potentially be probed by PTA collaborations~\cite{Franciolini:2023pbf,Firouzjahi:2023lzg,Liu:2023ymk,Balaji:2023ehk,Esmyol:2025ket,Gouttenoire:2025jxe,Bhaumik:2025vlb}. A particularly interesting research direction is therefore to study USR models capable of generating a detectable SIGW signal associated with heavy PBHs, while remaining consistent with CMB constraints. Alternatively, one may ask how current and future CMB measurements restrict the possible interpretation of a PTA signal as evidence for PBH-induced SIGWs. Finally, another natural extension is to incorporate additional observational constraints beyond those of the CMB, such as FIRAS limits on spectral distortions and Lyman-$\alpha$ forest data. Including these data sets would further reduce the available parameter space of USR models and sharpen the connection between PBH phenomenology, inflationary dynamics, and gravitational wave observations.

\vspace{-6mm}
\section*{Acknowledgments}
\vspace{-3mm}
We thank N. Bernal, G. Franciolini, G.Perna, A. Riotto, and K. Schmitz for constructive discussions.
This work was supported by the Estonian Research Council grants PSG869, RVTT3, RVTT7, TARISTU24-TK3, TARISTU24-TK10 and the Center of Excellence program TK202.

\bibliography{main}
\newpage

\onecolumngrid
\setcounter{equation}{0}
\setcounter{section}{0}
\setcounter{table}{0}
\setcounter{figure}{0}
\makeatletter
\renewcommand{\theequation}{A\arabic{equation}}
\renewcommand{\thefigure}{A\arabic{figure}}
\renewcommand{\thetable}{A\arabic{table}}
\appendix

\section{Slow-roll estimates}

\subsection{Slow roll beyond the leading order}
\label{app:SRandCR}

Even if numerical estimates are available, analytic approximations are useful as they can provide a deeper understanding of the results. Such approximations for the power spectra are available during the SR phase, when the CMB spectrum is created, and in the CR or SR phase that follows USR. SR power spectra beyond the leading order, and the CR power spectra, are given by~\cite{Stewart:1993bc,Gong:2001he},
\bea\label{eq:Pzeta_SR}
    \mathcal{P}_\zeta(k) 
    &= \left.\frac{\Gamma(\nu)^2}{(2\pi)^3 }\left( 2(1- \eps_1)\right)^{2\nu -1}\frac{H^2}{\MPl^2 \eps_1}\right|_{N=N_k}\,,
    \\
    \mathcal{P}_T(k) 
    &= \left.\frac{2\Gamma(\mu)^2}{\pi^3}\left( 2(1- \eps_1)\right)^{2\mu -1}\frac{H^2}{\MPl^2}\right|_{N=N_k}\,,
\eea
where $\mu \equiv 1/2 + 1/(1-\eps_1)$ and $\nu$ is given by Eq.~\eqref{eq:nu}. The SR case is recovered by setting $\nu = 3/2$. As we show in Section~\ref{sec:fit}, the polynomial USR models can exhibit a sizable $\eps_3$ at CMB scales and thus the second-order expansion is required to obtain sufficiently accurate estimates in SR. The inflationary parameters estimated at second-order in SR read~\cite{Gong:2001he}\footnote{The scalar spectral index $n_s$ and its running $\alpha_s$ are defined at a pivot scale $k_{\rm CMB} = 0.05 \,{\rm Mpc}^{-1}$ and the tensor-to-scalar ratio is calculated on the reference scale $k=0.002\,{\rm Mpc}^{-1}$.}
\bea\label{eq:ns,r_SR}
    r 
    &\equiv \frac{\mathcal{P}_T}{\mathcal{P}_\zeta}
    = 16\eps_1 \left(1+C(\eps_2-8 \eps_1/3) \right)+\mathcal{O}(\eps^3)\,, 
    \\
    n_{s}
    &\equiv 1 + \frac{\td \ln \mathcal{P}_\zeta}{\td \ln k}
    = 1-2\eps_1-\eps_2-2\eps^2_1-(3-2 C/3)\eps_1\eps_2-C\eps_2\eps_3+\mathcal{O}(\eps^3)\,,
    \\
    \alpha_s 
    &\equiv \frac{\td^2 \ln \mathcal{P}_\zeta}{\td \ln k^2}
    = -\eps_2(2\eps_1 + \eps_3)+\mathcal{O}(\eps^3)\,,
\eea
where $C \equiv \gamma+\ln2-2\simeq-0.73$ and $\gamma$ is the Euler constant.

Estimating the Hubble-flow parameters requires knowledge of background evolution, which, in USR models, must typically be obtained numerically. In SR, they can, however, be estimated directly from the potential using the potential SR parameters,
\bea\label{eq:pot_SR_params}
    \eps_V 
    &\equiv \frac{\MPl^2}{2} \left(\frac{V'}{V}\right)^2
    = \frac{\MPl^2}{2 \bar{K}}\left(\frac{\pd_\phi \bar V}{\bar V}\right)^2\,,
    \\
    \eta_V 
    &\equiv \MPl^2\frac{V''}{V}
    = \frac{\MPl^2}{\bar{K} \bar V} \left( \pd_\phi^2 \bar V-\frac{1}{2}\pd_\phi \bar V \pd_\phi \ln \bar{K} \right) \,,
    \\
    \xi_V^2 
    &\equiv \MPl^4 \frac{V''' V'}{\bar{K}^2V^2} 
    = \frac{\MPl^4 \pd_\phi \bar V}{\bar{K}^2V^2} \left( 
        \pd_\phi^3 \bar V 
    -   \frac{3\pd_\phi \bar{K}}{2\bar{K}}\pd_\phi^2 \bar V 
    -   \frac{\pd_\phi^2 \bar{K}}{2\bar{K}}\pd_\phi \bar V 
    +   \left(\frac{\pd_\phi \bar{K}}{\bar{K}}\right)^2\pd_\phi \bar V 
    \right)\,,
   \\ \sigma_V^3 &\equiv \MPl^5 \frac{V'''' V'^2}{V^3} =   \frac{\MPl^5 (\pd_\phi \bar V)^2}{\bar{K}^3V^3} \Bigg[  \pd^4_\phi \bar V -3 \frac{\bar K}{K} \pd_\phi^3 \bar V + \frac{19}{4} \left( \frac{\pd_\phi \bar K}{\bar K}\right)^2 \pd^2_\phi \bar V -2\frac{\pd_\phi^2 \bar K}{\bar K} \pd^2_\phi \bar V -  \\
   & \qquad \qquad \qquad \qquad \qquad \qquad \quad  \ - \frac{7}{2} \left( \frac{\pd_\phi \bar K}{\bar K}\right )^3 \pd_\phi \bar V + \frac{13\pd_\phi^2 \bar K \pd_\phi \bar K}{4 \bar{K}^2} \pd_\phi \bar V  - \frac{\pd_\phi^3 \bar K }{2 \bar K} \pd_\phi \bar V \Bigg ]
\eea
where $'$ denotes derivation with respect to the Einstein frame field $\bar\phi$ and then expressed here in terms of the Jordan frame field $\phi$. %During SR, the first two parameters are approximately $\eps_1 \approx \eps_V$, $\eps_2 \approx -2\eta_V + 4\eps_V$.

Using the second-order relation between $\eps_1$ and the potential SR parameters~\cite{Liddle:1994dx}
\be\label{eq:eps1V}
    \eps_1 = \eps_V - \frac{4}{3} \eps_V^2 + \frac{2}{3} \eps_V \eta_V + \mathcal{O}_3\,, \\
\ee
we can obtain the remaining Hubble flow parameters $\eps_i$ in terms of potential SR parameters by applying the definition of $\eps_i$ and using $\pd_N = y \pd_\phi$ when taking derivatives of the potential. At next-to-leading order, we have
\bea\label{eq:epsi_epsV}
    \eps_2 &= 4 \eps_V - 2 \eta_V - 8 \eps_V^2- \frac{2}{3} \eta_V^2 + \frac{20}{3} \eps_V \eta_V -\frac{2}{3} \xi^2_V + \mathcal{O}_3\,,\\
    \eps_3 &= -\eps_2^{-1}\left(
    -16\eps_V^2 + 12 \eps_V \eta_V - 2\xi_V^2 
    + \frac{224}{3} \eps_V^3 - \frac{256}{3} \eps_V^2 \eta_V+20 \eps_V \eta_V^2 + \frac{32}{3}\eps_V \xi_V^2 - \frac{8}{3}\eta_V \xi_V^2 - \frac{2}{3}\sigma_V^3
    \right) + \mathcal{O}_3\,,
\eea
We didn't expand the $\eps_2^{-1}$ prefactor in the expansion of $\eps_3$, because it can make the approximation fail when $\xi_V^2 \gg \eps_V, \eta_V$. This expansion agrees with Ref.~\cite{Liddle:1994dx}, where the relations were given in terms of the Hubble SR parameters $\eta_H \equiv 2\MPl^2 H''/H$ and $\xi_H^2 \equiv 2\MPl^4 H'''H'/H^2$, where $'$ denotes derivation with respect to the field in the Einstein frame~\cite{Liddle:1994dx}. They are related to the Hubble flow parameters $\eps_i$ by $\eps_2 = 2(\eps_1-\eta_H)$ and $\eps_3 = (2\eps_H^2 - 3\eps_1\eta_H + \xi_H^2)/(\eps_1-\eta_H)$.

\subsection{Inflection point inflation in slow-roll}
\label{app:IPI}

\begin{figure*}[t!]
	\centering
\includegraphics[width=0.6\textwidth]{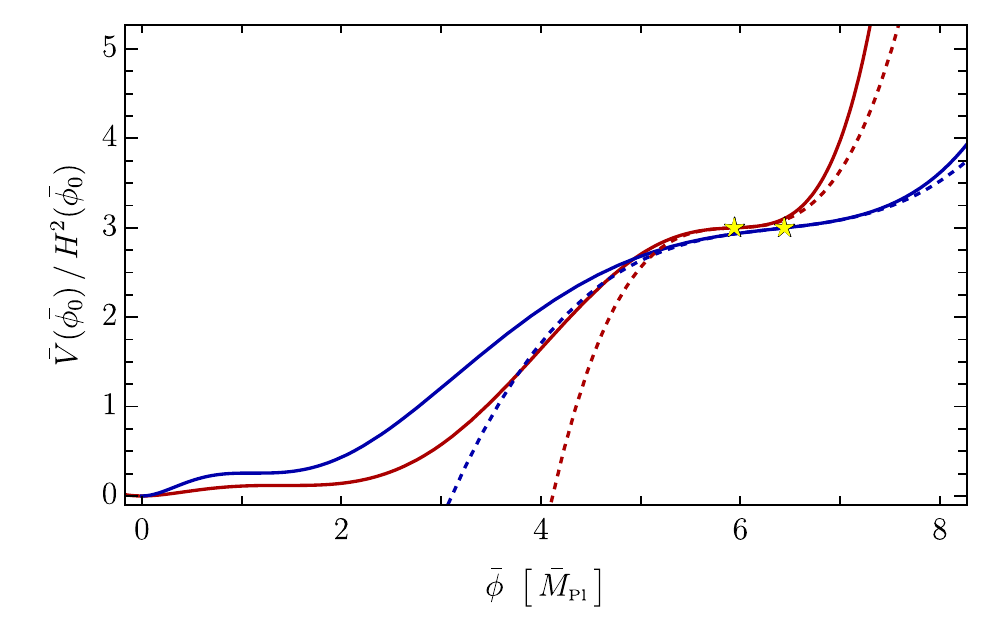}
	\caption{ \justifying  \it Comparison between the Einstein-frame potentials of two benchmark realizations of the NMC (blue) and MC (red) models and their corresponding fourth-order expansions (dashed lines) around the inflection point, as defined in Eq.\,\eqref{eq:quasi_inflection_point}. The yellow stars indicate the positions of the inflection points.}
\label{fig:FitCMB}
\end{figure*}

As is clear from Fig.\,\ref{fig:FitCMB}, both the NMC and MC models can be approximated by the expansion in Eq.\,\eqref{eq:quasi_inflection_point} around the CMB scales.
This gives us the possibility to estimate $n_s$, $\alpha_s$ and $r$ directly from the cubic-order potential, reducing the parameter space to $V_0,\,\beta$ and $\delta$.
Given the potential of Eq.\,\eqref{eq:quasi_inflection_point} and the definitions of Eqs.\eqref{eq:pot_SR_params}, the SR parameters can be easily computed analytically:
\bea \label{eq:newEpsEtaXi}
    \eps_V = \frac{  (\beta + 3 \delta^2)^2}{ 2 \bar{V}^2(\delta)}\,,
    \qquad
    \eta_V = \frac{6 \delta}{ \bar{V} (\delta)}\,, 
    \qquad
    \xi_V^2 \simeq  \frac{6 (\beta + 3 \delta^2)}{ \bar{V}^2 (\delta)}\,, \qquad \sigma_V^3 = 0 \,.
\eea

Since this expansion around the inflection point is intended to parametrize the behavior of our PBH-producing models during the initial SR regime, we can compute $N_\textrm{SR} \equiv N_1$ as the number of e-folds from the horizon exit of the reference scale, $y_\star$, up to the breakdown of slow roll.
The number of e-folds at the lowest order can be expressed as
\bea\label{eq:N1_app}
     N_1 (\delta_\star)
    &\simeq N_0 + \int_{0}^{\delta_\star} \frac{1}{\sqrt{2 \eps_V}} d\delta
    \\
    &= \frac{V_0}{\sqrt{3 \beta}} \textrm{atan}\left(\frac{\sqrt{3}\delta}{\sqrt{\beta}}\right) + \frac{\delta^2}{6} + \frac{\beta}{9}\ln(\beta + 3x^2) + N_0
    \\
    &\simeq \frac{V_0}{\sqrt{3 \beta}} \textrm{atan}\left(\frac{\sqrt{3}\delta}{\sqrt{\beta}}\right) + N_0
\eea
where $N_0$ quantifies the time it takes to roll from the inflection point, i.e., $\delta=0$, to the breakdown of the first SR phase. It depends on the model parameters but not on the field $\delta$. As the potential at low field values is not captured well by the cubic expansion, we will fix $N_0$ from the numerical solution of the exact field equation. Inverting \eqref{eq:N1_app} gives
\eqref{eq:N1_IP}.

Replacing the analytical expressions of the potential SR parameters from Eq.\,\eqref{eq:newEpsEtaXi} into Eqs.\,\eqref{eq:eps1V} and \eqref{eq:epsi_epsV}, we can compute the Hubble parameters $\eps_i$ directly from the potential.
We stress that, upon expanding the NMC and MC models under consideration, we found that $0 < \beta \lesssim 1$ with $\beta, |\delta| \ll V_0$, and that around the CMB scales $|\delta| \ll 1$ by construction. The Hubble SR parameters at the next-to-leading order in $\beta$, $\delta$ and $1/V_0$ are given by
\be \label{eq:LOepsi}  
    \eps_1 \simeq \frac{\beta^2}{2 V_0^2} + \frac{48\beta \delta^2}{V_0^2}\,, \qquad 
    \eps_2 \simeq -\frac{12 \delta}{V_0} - \frac{4 \beta}{V_0^2}\,, \qquad
    \eps_3 \simeq -\frac{3 \beta}{3 V_0 \delta + \beta}\,,   
\ee
where the denominator of $\eps_3$ should not be expanded. Eq.~\eqref{eq:ns,r_SR} then gives the inflationary observables at the second order in SR. Therefore, to capture the evolution of inflationary parameters, we can expand them $\beta$ and $y$ to obtain Eq.~\eqref{eq:SRexpansion}.

\subsection{The log+bump toy model}
\label{app:toyexpl}
Here we report the full results for the Hubble and SR parameters (at the first order) for the logarithmic toy model~\eqref{eq:V_log+bump} studied in Sec.~\ref{sec:modelbulding}. The first potential SR parameter away from the Gaussian feature is
\be
    \eps_V = \frac{\MPl^2 \beta^2}{2\phi^2 \left(1 + \beta \ln\left( \phi/\MPl \right)\right)^2}\,.
\ee
The number of $e$-folds can then be estimated from
\be
    N 
    = \int \frac{\td \phi}{\sqrt{2\eps_V}} 
    = \frac{\phi^2}{4\beta \MPl^2}\left(2 - \beta + 2\beta \ln\left( \phi/\MPl \right) \right) - N_{0}
\ee
where $N_{0}$ is a constant determined by setting the end of inflation $\eps_V = 1$ to $N=0$. It is given by
\be\label{eq:toy_N0}
    N_{0} = \frac{2W\left(e^{1/\beta}/\sqrt{2}\right)-1}{8W\left(e^{1/\beta}/\sqrt{2}\right)^2}\,.
\ee
so that
\be\label{eq:toy_phiN}
    \phi(N) = 2\sqrt{\frac{N+N_0}{W_N}}\,,
\ee
where $W_N \equiv W\Big(4(N+N_{0})\,e^{2/\beta-1}\Big)$. By substituting \eqref{eq:toy_phiN} into the potential SR parameters, we then obtain
\bea\label{eq:eps_i_toy}
    \eps_1(N) &= \frac{W_N}{2\,(N+N_{0})\,(1+W_N)^2}, \\
    \eps_2(N) &= \frac{W_N\,(W_N+3)}{(N+N_{0})\,(1+W_N)^2}, \\
    \eps_3(N) &= \frac{1}{(N+N_{\rm end})} \left[ 1 - \frac{1}{1+W_N} - \frac{W_N}{(1+W_N)(W_N+3)} + \frac{2 W_N}{(1+W_N)^2} \right]\,.
\eea
The $\beta\to0$ limit yields Eq.~\eqref{eq:app1}. 

We note that since $\eps_V$ is growing monotonically, we can invert the expression for $\eps_V(N)$,
\be
    N(\eps_V) 
    = \frac{2 W\left(e^{1/\beta}/\sqrt{2 \eps_V}\right) - 1}{8 \eps_V W\left(e^{1/\beta}/\sqrt{2\eps_V}\right)^2} - N_0\,,
\ee
where $N_{0}$ is then obtained by setting $\eps_V=1$. This inversion was used to obtain \eqref{eq:toy_N0}. Furthermore, when $\beta \ll 1$, $N_{0} \sim \beta/4$, so $N_{0} \approx 0$ is a reasonable approximation in that limit.

\renewcommand{\theequation}{B\arabic{equation}}
\renewcommand{\thefigure}{B\arabic{figure}}
\renewcommand{\thetable}{B\arabic{table}}
\section{Full MCMC results for the NMC and MC models.}
\label{app1}

The full MCMC posterior distributions obtained for both the MC and NMC cases are reported in this Appendix. It is important to note that the following discussion regarding the MCMC analysis should not be interpreted as a \textit{classical statistical} inference, rather as an algorithmic exploration of the parameter space of our inflationary models, with the primary goal of identifying regions that are potentially consistent with current observational constraints.

It is well known that PBH models require a degree of tuning in order to generate a sizable peak in the power spectrum~\cite{Cole:2023wyx,Karam:2023haj}. To overcome the related issues, we fixed the starting points of the Markov chains by hand to guarantee a spectral peak. 
The MCMC scans were performed using the Python package \texttt{emcee}~\cite{Foreman-Mackey:2012any}, which implements an affine-invariant ensemble sampler.
We use 32 walkers initialized around the starting points listed in Tab.\,\ref{tab:injected}. 

For the MC models, we found that scanning the parameter space using 
the parametrization with an explicit double inflection point~\cite{Allegrini:2024ooy} 
results in a much faster convergence of the MCMC. 
For completeness, we report the parametrization:
\bea
\bar{V}(x) = a_4 \MPl^4
\Bigg\{
x^4 + \frac{2}{x_0^2+4x_0x_1 + x_1^2} \Bigg[ & 
x_0^2x_1^2(1+\beta_2)x^2 -
\frac{4}{3}x_0x_1(x_0+x_1)(1+\beta_3)x^3 \\
 & -
\frac{4}{5}(x_0 + x_1)(1+\beta_5)x^5 + \frac{(1+\beta_6)x^6}{3}
\Bigg]
\Bigg\}\,,\label{eq:DoubleInflectPote}
\eea
here $x_0,\,x_1$ denote the positions of the two inflection points, and the $\beta_i$ parameters represent deviations from the stationary case.
The results were then expressed in terms of $\tilde a_i$ to enable a direct comparison with the NMC models.
    
\begin{table}[htbp]
    \centering

  % --------- Tabella NMC ---------
    \begin{minipage}{0.99\textwidth}
        \centering
        \begin{tabular}{|wc{10mm}|wc{1.8cm} wc{1.8cm} wc{1.8cm} wc{1.8cm} wc{1.8cm} wc{1.8cm} |wc{3cm}|} 
            \hline
            &$\xi$ &$ \tilde{a}_2 $& $\tilde{a}_3$ & $\tilde{a}_5\ \ [10^{-2}]$ &
            $\tilde{a}_6 \ \ [10^{-3}]$ &
            $\xi_1$ & $M_H (k_\textbf{pk})/M_\odot$\\
            \hline
            \multirow{3}{*}{\rotatebox[origin=c]{90}{\bf{NMC}}}
            & $0.2772$ & $2.173$ & $-2.494$ & $-6.23$  & $2.23$ & $0.0$  & $4.0 \times10^{-1}$
            \\ 
            & $0.2678$ & $2.158$ & $-2.494$ & $-6.23$ & $2.07$ & $0.0$ & $4.0\times10^{-6}$
            \\ 
            & $0.2491$ & $2.126$ &$-2.494$ &  $-6.23$ & $1.97$ & $0.0$ & $2.5 \times 10^{-12}$ \\
            %& $\phi_0\ [\bar{M}_{\rm Pl}]$ & $10.74$ & $10.8$ & $10.93$
            
            \hline%[dashed]
            \multirow{3}{*}{\rotatebox[origin=c]{90}{\bf{MC}}}
           & $0.0$ & $2.015$ & $-2.338$ & $-16.54$  & $9.35$ & $0.0$  & $1.0 \times10^0$
            \\ 
            & $0.0$ & $1.971$ &$-2.313$ &  $-16.68$ & $9.51$ & $0.0$ & $6.3\times 10^{-4}$
            \\
            & $0.0$ & $1.783$ &$-2.201$ & $-17.57$ & $10.55$ & $0.0$ & $4.0 \times10^{-14}$ \\
            %& $\phi_0\ [\bar{M}_{\rm Pl}]$ & $6.04$ & $5.94$ & $5.82$
            
            \hline
        \end{tabular} 
    \end{minipage}

\caption{ The list of starting points of the MCMC chains. The MC scenarios use chains for which $\xi = 0$. In the rightmost column, we report the horizon mass at the peak of the numerical power spectrum.}
    \label{tab:injected}
    \vspace{-0.0cm}
\end{table}

The \textit{mock}-likelihood is constructed to have an efficient exploration of the parameter space, potentially ending up with configurations in agreement with the latest cosmological constraints.
In particular, four MCMC analyzes were performed to explore polynomial inflation, in regions consistent with the Planck~\cite{Planck:2018vyg} and ACT+Planck~\cite{ACT:2025tim} measurements of $r$, $n_s$ and $\alpha_s$. 
The \textit{mock}-likelihood employed in the code reads
\bea
    \mathcal{L}(D) 
    = 
    \mathcal{L}_{\rm CMB}(D\, | \boldsymbol{\theta} ) 
    \times \mathcal{L}_{\rm PBH}(D\, | \,\mathcal{P}_\mathcal{\zeta}(k_{\rm pk})) \times \mathcal{L}(D\, | \,N_1) \,, 
\eea
where for brevity $\boldsymbol{\theta} = \{\,n_s,\,\alpha_s,\,r,\, \mathcal{P}_{\zeta}(k_{\rm pk}),\,N_1\}$.
\be
     \mathcal{L}_{\rm CMB}(D\, | \boldsymbol{\theta} ) 
     = \exp\left( - \frac{1}{2} K^{-1}_{ij} \left(\theta^{\rm CMB}_i(D) - \bar\theta^{\rm CMB}_{i}\right)\left(\theta^{\rm CMB}_j(D) - \bar\theta^{\rm CMB}_{j}\right) \right)
\ee 
represents a successful realization of an inflationary model defined by its parameters 
$D = \{\, \xi_1,\, \xi,\, \tilde a_2,\, \tilde a_3,\, \tilde a_5,\, \tilde a_6 \,\}$. 
The CMB observables are modeled using Gaussian distributions centered at 
$\bar\theta^{\rm CMB}_{i}$ and assuming a diagonal covariance matrix 
$K_{ij} = \delta_{ij}\,\sigma^2(\bar \theta^{\rm CMB}_{i})$. 
The values of $\bar\theta^{\rm CMB}_{i}$ are chosen depending on whether 
the MCMC is designed to explore the Planck or ACT+Planck allowed regions. 
For the corresponding central values and uncertainties, we refer the reader 
to Sec.\,\ref{sec:Introduction}. 
The remaining part of $\mathcal{L}(D)$ is parametrized as follows:
\bea
    \mathcal{L}_{\rm PBH}(D\, | \,\mathcal{P}_\mathcal{\zeta}(k_{\rm pk}) ) & = \exp\left(-\frac{(\log_{10}\mathcal{P}_\mathcal{\zeta}(k_{\rm pk}) - \log_{10}\bar{\mathcal{P}}_\mathcal{\zeta})^2}{2 \sigma^2_{\bar{\mathcal{P}}_\mathcal{\zeta}}}\right)\,, \\ \mathcal{L}(D\, | \,N_1) & = 
\frac{1}{(1 + e^{-10 (N_1 - N_1^\textrm{max})})(1 + e^{10 (N_1 - N_1^\textrm{min})})}\,,
\eea
where $\mathcal{L}_{\rm PBH}(D)$ is the probability that the parametrization $D$ yields a sizable PBH abundance, 
with $(\log_{10}\bar{\mathcal{P}}_{\zeta},\,\sigma_{\bar{\mathcal{P}}_{\zeta}})=(-3.2,\,0.2)$. 
The window-function $\mathcal{L}(D\mid N_1)$ constrains the scanned mass range via Eq.\,\eqref{eq:kN1}, 
with $(N_1^{\rm min},\,N_1^{\rm max})=(17,\,37)$. 
For modeling $\mathcal{L}_{\rm PBH}$ we considered the difference in the peak height between the SR and numerical estimates of the power spectrum.

To produce well-behaved sampling weights for plotting and comparison, we rescale the original per-sample likelihood estimates. These steps are heuristic and should not be interpreted as a rigorous recalibration of observational likelihoods. For this reason, we refer to our estimate as a \textit{mock}-likelihood.
Given this, the prior for each configuration of the models can be computed as
\be
    p(\boldsymbol{\theta} \,|\, D) \propto \pi(\boldsymbol{\theta})\,\tilde{\mathcal{L}} (D \,|\, \boldsymbol{\theta})\,,
\ee
where $\tilde{\mathcal{L}}$ denotes the renormalized likelihood employed in the code and $\pi(\boldsymbol{\theta})$ denotes some priors for the parameters. 
Flat priors were assumed in our scans.
Due to the severe fine-tuning required to enhance the peak and to efficiently explore the parameter space, the acceptance rate ($A$) of the MCMC was modified to lie in the range $A \in [0.01,\, 1]$ for $\mathcal{L}(D \,|\, \boldsymbol{\theta}) \in [10^{-10},\, 1]$. 

\begin{figure*}[t!]
	\centering
\includegraphics[width=0.95\textwidth]{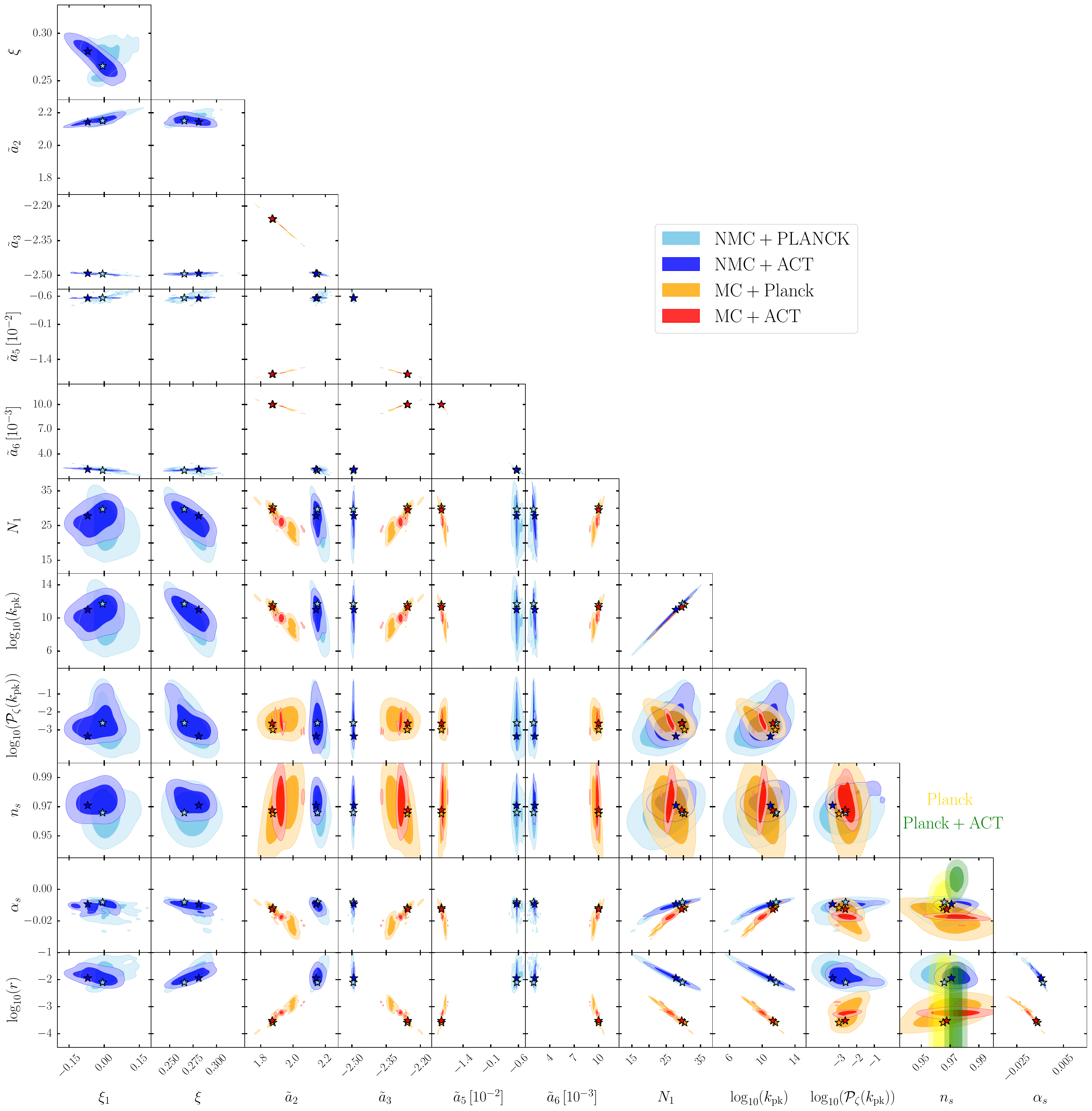}
	\caption{ \justifying  \it Corner plot posteriors for the full parameter space of the NMC and MC models, using the injected potential defined in Tab.~\ref{tab:injected}. The coloured stars highlight the best-fit points from the MCMC runs using two different Gaussian likelihoods, centered respectively on Planck data and Planck+ACT as described in App.\,\ref{app1}.
 }
\label{fig:NMCCorner}
\end{figure*}

Fig.~\ref{fig:NMCCorner} shows the posterior distributions for the NMC and MC models, respectively, obtained from the different MCMCs. The desaturated colors correspond to Planck likelihood analyses, whereas the saturated ones indicate Planck+ACT likelihood analyses.

To speed up the scan, the inflationary observables $(n_s,\,\alpha_s,\,r)$, together with the peak amplitude of the power spectrum $\mathcal{P}_{\zeta}(k_{\rm pk})$, were first computed using the second-order SR approximations~\eqref{eq:Pzeta_SR} and~\eqref{eq:ns,r_SR}, and then recomputed by solving the Mukhanov–Sasaki equation~\eqref{eq:MukSas} for all points shown in Fig.~\ref{fig:Corner} and Fig.~\ref{fig:NMCCorner}. As a side product, we confirmed that for $(n_s,\,\alpha_s,\,r)$ the second-order SR approximation was in good agreement with the full numerical solutions of the Mukhanov-Sasaki equation~\eqref{eq:MukSas}.

\end{document}